%% file: main.tex
\documentclass[review,12pt]{elsarticle}

\makeatletter
\def\ps@pprintTitle{%
 \let\@oddhead\@empty
 \let\@evenhead\@empty
 \def\@oddfoot{\centerline{\thepage}}%
 \let\@evenfoot\@oddfoot}
\makeatother

\usepackage{amsmath}
\usepackage{relsize}
\usepackage{amssymb}
\usepackage{lineno}
\usepackage{color}
\usepackage{graphicx}
\usepackage{enumitem}

\usepackage{caption}
\usepackage{enumitem}
\usepackage{csvsimple}
\graphicspath{ {Figures/} }
\usepackage{booktabs}
\usepackage{tabulary}
\usepackage[table,xcdraw]{xcolor}
\usepackage{subcaption}
\usepackage{lscape}
\usepackage[margin=1in]{geometry}
\usepackage{hyperref}
\usepackage{url}
\usepackage{framed}
\usepackage{pdfpages}
\usepackage{algorithm}
\usepackage[noend]{algpseudocode}
\usepackage{mathrsfs}
\usepackage{mathtools}
\usepackage{soul}
\usepackage[export]{adjustbox}
\usepackage{longtable,lscape}
\usepackage[export]{adjustbox}
\usepackage[section]{placeins}

\journal{TBD}
 \usepackage{color}

\begin{document}

\begin{frontmatter}

\title{Stock2Vec: A Hybrid Deep Learning Framework for \\Stock Market Prediction with Representation Learning \\and Temporal Convolutional Network}

\author[a]{Xing Wang}
\author[b]{Yijun Wang}
\author[c]{Bin Weng}
\author[a]{Aleksandr Vinel}
\address[a]{Department of Industrial Engineering, Auburn University, AL, 36849, USA}
\address[b]{Verizon Media Group (Yahoo!), Champaign, IL, 61820, USA}
\address[c]{Amazon.com Inc., Seattle, WA, 98108, USA}

\begin{abstract}
\input{stock/abstract}
\end{abstract}


\end{frontmatter}

\section{Introduction}
\label{sec:s2v_intro}
\input{stock/intro}

\section{Related Work}
\label{sec:s2v_bg}
\input{stock/related}
\section{Methodology}
\label{sec:s2v_method}
\input{stock/method}

\section{Data Specification}
\label{sec:s2v_data}
\input{stock/data}

\section{Experimental Results and Discussions}
\label{sec:s2v_res}
\input{stock/results}

\section{Concluded Remarks and Future Work}
\label{sec:s2v_conc}
\input{stock/conclusions}

\linespread{1} \selectfont
\label{sec:ref}
 \bibliographystyle{elsarticle-num}
\bibliography{main.bbl}

\FloatBarrier
\appendix

\section{Sector Level Performance Comparison}
\label{sec:s2v_sectorTab}
\input{stock/tabs/sectorRMSE}
\input{stock/tabs/sectorMAE}
\input{stock/tabs/sectorMAPE}
\input{stock/tabs/sectorRMSPE}

\FloatBarrier
\newpage
\section{Performance comparison of different models for the one-day ahead forecasting on different symbols}
\label{sec:s2v_symbTab}
\input{stock/tabs/s20RMSE}

\input{stock/tabs/s20MAE}
\input{stock/tabs/s20MAPE}
\input{stock/tabs/s20RMSPE}

\FloatBarrier
\section{Plots of the actual versus predicted prices of different models on the test data}
\label{sec:s2v_plots21}
\input{stock/plots21}

\end{document}

%% file: stock/abstract.tex
We have proposed to develop a global hybrid deep learning framework to predict the daily prices in the stock market. With representation learning, we derived an embedding called Stock2Vec, which gives us insight for the relationship among different stocks, while the temporal convolutional layers are used for automatically capturing effective temporal patterns both within and across series. Evaluated on S\&P 500, our hybrid framework integrates both advantages and achieves better performance on the stock price prediction task than several popular benchmarked models.


%% file: stock/intro.tex
In  finance, the classic strong efficient market hypothesis (EMH) posits that the stock prices follow random walk and cannot be predicted \cite{fama1965behavior}. Consequently, the well-known capital assets pricing model (CAPM) \cite{sharpe1964capital, lintner1975valuation, black1972capital} serves as the foundation for portfolio management, asset pricing, among many applications in financial engineering. The CAPM assumes a linear relationship between the expected return of an asset (e.g., a portfolio, an index, or a single stock) and its covariance with the market return, i.e., for a single stock, CAPM simply predicts its return $r_i$ within a certain market with the linear equation $$r_i(t) = \alpha_i + \beta_i r_{m}(t),$$
where the Alpha ($\alpha_i$) describes the stock's ability to beat the market, also refers to as its ``excess return'' or ``edge'', and the Beta ($\beta_i$) is the sensitivity of the expected returns of the stock to the expected market returns ($r_m$). Both Alpha and Beta are often fitted using simple linear regression based on the historical data of returns. With the efficient market hypothesis (EMH), the Alphas are entirely random with expected value of zero, and can not be predicted. 

In practice, however, financial markets are  more complicated than the idealized and simplified strong EMH and CAPM. 
Active traders and empirical studies suggest  that the financial market is never perfectly efficient and thus the stock prices as well as the Alphas can be predicted, at least to some extent. Based on this belief, stock prediction has long played a key role in numerous data-driven decision-making scenarios in financial market, such as deriving trading strategies, etc. 
Among various methods for stock market prediction, the classical Box-Jenkins models \cite{box1968some}, exponential smoothing techniques, and state space models \cite{hyndman2008forecasting} for time series analysis are most widely adopted, in which the factors of autoregressive structure, trend, seasonality, etc. are independently estimated from the historical observations of each single series. 
In recent years, researchers as well as the industry have deployed various machine learning models to forecast the stock market, 
such as k-nearest neighbors (kNN) \cite{alkhatib2013stock, chen2017feature}, hidden Markove model (HMM) \cite{hassan2007fusion, hassan2013hmm}, 
support vector machine (SVM) \cite{yang2002support, huang2009hybrid}, 
artificial neural network (ANN) \cite{wang2011forecasting, guresen2011using,  kristjanpoller2014volatility, wang2015back, goccken2016integrating}, 
and various hybrid and ensemble methods \cite{patel2015predicting, booth2014automated, barak2015developing, patel2015predicting,  weng2018macroeconomic}, 
among many others.
The literature has demonstrated that machine learning models typically outperform traditional statistical time series models, which might be mainly due to the following reasons: 1) less strict assumption for the data distribution requirement, 2) various model architecture can effectively learn complex linear and non-liner from data, 3) sophisticated regularization techniques and feature selection procedures provide flexibility and strength in handling correlated input features and control of overfitting, so that more features can be thrown in the machine learning models.
As the fluctuation of the stock market indeed depends on a variety of related factors,
in addition to utilizing the historical information of stock prices and volumes as in traditional technical analysis \cite{murphy1999technical}, 
recent research of stock market forecasting has been focusing on informative external source of data, 
for instance, the accounting performance of the company \cite{fama1993common}, macroeconomic effects \cite{tetlock2008more, weng2018macroeconomic}, government intervention and political events \cite{li2016tensor}, etc. 
With the increased popularity of web technologies and their continued evolution, the opinions of public from relevant news \cite{weng2018predicting} and social media texts \cite{bollen2011twitter, oliveira2017impact} have an increasing effect on the stock movement, 
various studies have confirmed that combining the extensive crowd-sourcing and/or financial news data facilitates more accurate prediction \cite{wang2018combining}. 

During the last decade, with the emergence of deep learning, various neural network models have been developed and achieved success in a broad range of domains, such as computer vision \cite{lecun1998gradient, krizhevsky2012imagenet, simonyan2014very, redmon2016yolo} and natural language processing
\cite{mikolov2013efficient, pennington2014glove, devlin2018bert}. 
For stock prediction specifically, recurrent neural networks (RNNs) are the most preferred deep learning models to be implemented \cite{rather2015recurrent, fischer2018deep}. 
Convolutional neural networks (CNNs) have also been utilized, however, most of the work transformed the financial data into images to apply 2D convolutions as in standard computer vision applications. For example, the authors of 
\cite{sezer2018algorithmic} converted the technical indicators data to 2D images and classified the images with CNN to predict the trading signals. Alternatively, 
\cite{hu2018candlestick} directly used the candlestick chart graphs as inputs to determine the Buy, Hold and Sell behavior as a classification task, while in \cite{sezer2019bar}, the bar chart images were fed into CNN.  The authors of
\cite{hoseinzade2019cnnpred} uses a 3D CNN-based framework to extract various sources of data including different markets for predicting the next day's direction of movement of five major stock indices, which showed a significant improved prediction performance compared to the baseline algorithms.
There  also exists research  combining RNN and CNN together, in which the temporal patterns were learned by RNNs, while CNNs were only used for either capturing the correlation between nearby series (in which the order matters if there are more than 2 series) or learning from images, see \cite{long2019deep, jiang2017deep}. Deployment of CNN in all these studies differs significantly from ours, since we aim at capturing the temporal patterns without relying on two-dimensional convolutions. In \cite{di2016artificial}, 1D causal CNN was used for making predictions based on the history of closing prices only, while no other features were considered. 


Note that all of the aforementioned work has put their effort into learning more accurate Alphas, and most of the existing research focuses on deriving separate models for each of the stock, while only few authors consider the correlation among different stocks over the entire markets as a possible source of information. 
In other words, the Betas are often ignored. At the same time, since it is natural to assume that markets can have nontrivial correlation structure, it should be possible to extract useful information from group behavior of assets. 
Moreover, rather than the simplified linearity assumed in CAPM, the true Betas may exhibit more complicated nonlinear relationships between the stock and the market. 

In this paper, we propose a new deep learning framework that leverages both the underlying Alphas and (nonlinear) Betas. In particular, our approach innovates in the following aspects:
\begin{enumerate}[label=\arabic*)]
      \item from model architecture perspective, we build a hybrid model that combines the advantages of both representation learning and deep networks. With representation learning, specifically, we use embedding in the deep learning model to derive implicit Betas, which we refer to as Stock2Vec, that not only gives us insight into the correlation structure among stocks, but  also helps the model more effectively learn from the features thus improving prediction performance. In addition, with recent advances on deep learning architecture, in particular the temporal convolutional network, we further refine Alphas by letting the model automatically extract temporal information from raw historical series. 
    \item and from data source perspective, unlike many time series forecasting work that directly learn from raw series, we generate technical indicators features supplemented with  external sources of information such as online news. Our approach differs from most research built on machine learning models, since in addition to explicit hand-engineered temporal features, we use the raw series as augmented data input. More importantly, instead of training separate models on each single asset as in most stock market prediction research, we learn a global model on the available data over the entire market, so that the relationship among different stocks can be revealed.

\end{enumerate}

The rest of this paper is organized as follows. 
Section \ref{sec:s2v_bg} lists several recent advances that are related to our method, in particular deep learning and its applications in forecasting as well as the representation learning.
Section \ref{sec:s2v_method} illustrates the building blocks and details of our proposed framework, specifically, Stock2Vec embedding and the temporal convolutional network, as well as how our hybrid models are built.
Our models are evaluated on the S\&P 500 stock price data and benchmarked with several others, the evaluation results as well as the interpretation of Stock2Vec are shown in Section \ref{sec:s2v_res}. Finally, we conclude our findings and discuss the meaningful future work directions in Section \ref{sec:s2v_conc}.

%% file: stock/related.tex
Recurrent neural network (RNN) and its variants of sequence to sequence (Seq2Seq) framework \cite{sutskever2014sequence} have achieved great success in many sequential modeling tasks, such as machine translation \cite{cho2014learning}, speech recognition \cite{sak2014long}, natural language processing \cite{bengio2003neural}, and extensions to autoregressive time series forecasting \cite{salinas2019deepar, rangapuram2018deep} in recent years. However, RNNs can suffer from several major challenges. Due to its inherent temporal nature (i.e., the hidden state is propagated through time), the training cannot be parallelized. Moreover, trained with backpropagation through time (BPTT) \cite{werbos1990backpropagation}, RNNs severely suffer from the problem of gradient vanishing thus often cannot capture long time dependency \cite{pascanu2013difficulty}. More elaborate architectures of RNNs use gating mechanisms to alleviate the gradient vanishing problem, with
the long short-term memory (LSTM) \cite{hochreiter1997lstm} and its simplified variant, the gated recurrent unit (GRU) \cite{chung2014gru} being the two canonical architectures commonly used in practice.

Another approach, convolutional neural networks (CNNs) \cite{lecun1989backpropagation}, can be easily parallelized, and recent advances effectively eliminate the vanishing gradient issue and hence help building very deep CNNs. These works include the residual network (ResNet) \cite{he2016deep} and its variants such as highway network \cite{srivastava2015training}, DenseNet \cite{huang2017densely}, etc.
In the area of sequential modeling, 1D convolutional networks offered an alternative to RNNs for decades \cite{waibel1989phoneme}. 
In recent years, \cite{oord2016wavenet} proposed WaveNet, a dilated causal convolutional network as an autoregressive generative model. Ever since, multiple research efforts have shown that with a few modifications, certain convolutional architectures achieve state-of-the-art performance in the fields of audio synthesis \cite{oord2016wavenet},
language modeling \cite{dauphin2017language}, 
machine translation \cite{gehring2017convolutional},
action detection \cite{lea2017temporal},
and time series forecasting \cite{binkowski2018autoregressive, chen2020probabilistic}. 
In particular, \cite{bai2018empirical} abandoned the gating mechnism in WaveNet and proposed temporal convolutional network (TCN). The authors benchmarked TCN with LSTM and GRU on several sequence modeling problems, and demonstrated that TCN exhibits substantially longer memory and achieves better performance.

Learning of the distributed representation has also been extensively studied \cite{bengio2000modeling, paccanaro2001learning, hinton1986learning} with arguably the most well-known application being  word embedding \cite{bengio2003neural, mikolov2013efficient, pennington2014glove} in language modeling. Word embedding maps words and phrases into  distributed vectors in a semantic space in which words with similar meaning are closer, and some interesting relations among words can be revealed, such as
\begin{align*}
    \text{King} - \text{Man} & \approx \text{Queen} - \text{Woman} \\
    \text{Paris} - \text{France} & \approx \text{Rome} - \text{Italy}
\end{align*}
as shown in \cite{mikolov2013efficient}.
Motivated by Word2Vec, the neural embedding methods have been extended to other domains in recent years. The authors of  \cite{barkan2016item2vec} obtained item embedding for recommendation systems through a collaborative filtering neural model, and`   called it Item2Vec which is capable of inferring relations between items even when user information is not available. 
Similarly, \cite{choi2016multi} proposed Med2Vec that learns the medical concepts with the sequential order and co-occurrence of the concept codes within patients' visit, and showed higher prediction accuracy in clinical applications. 
In \cite{guo2016entity}, the authors mapped every categorical features into ``entity embedding'' space for structured data and applied it successfully in a Kaggle competition, they also showcased the learned geometric embedding coincides with the real map surprisingly well when projected to 2D space. 

In the field of stock prediction, 
the term ``Stock2Vec'' has  already been used before. Specifically,
\cite{minh2018deep} trained word embedding that specializes in sentiment analysis over the original Glove and Word2Vec language models, and using such a ``Stock2Vec'' embedding and a two-stream GRU model to generate the input data from financial news and stock prices, the authors predicted the price direction of S\&P500 index. The authors of
\cite{wu2019deep} proposed another ``Stock2Vec'' which also can be seen as a specialized Word2Vec, trained 
using the co-occurences matrix with the number of the news articles that mention both stocks as entries. 
 Stock2Vec model proposed here  differs from these homonymic approaches and has its distinct characteristics.
First, our Stock2Vec is an entity embedding that represent the stock entities rather than a word embedding that denotes the stock names with language modeling. As the difference between entity embedding and word embedding may seem ambiguous, more importantly, instead of training the linguistic models with the co-occurrences of the words, 
our Stock2Vec embedding is trained directly as features through the overall predictive model, with the direct objective that minimizes prediction errors, thus illustrating the relationships among entities, while the others are actually fine-tuned subset of the original Word2Vec language model.
Particularly inspiring for our work are the entity embedding \cite{guo2016entity} and the temporal convolutional network \cite{bai2018empirical}. 

%% file: stock/method.tex
\subsection{Problem Formulation}
We focus on  predicting the future values of  stock market assets given the past. More formally speaking,  our input consists of a fully observable time series signals $\mathbf{y}_{1:T} = (y_1, \cdots, y_T)$ together with another related multivariate series $\mathbf{X}_{1:T} = (\mathbf{x}_1, \cdots, \mathbf{x}_{T})$, in which $\mathbf{x}_t \in \mathbb{R}^{n-1}$, and $n$ is the total number of series in the data. We aim at generating the corresponding target series $\hat{\mathbf{y}}_{T+1:T+h} = (\hat{y} _{T+1}, \cdots, \hat{y}_{T+h}) \in \mathbb{R}^{h}$ as the output, where $h \ge 1$ is the prediction horizon in the future. To achieve the goal, we will learn a sequence modeling network with parameters $\theta$ to obtain a nonlinear mapping from the input state space to the predicted series, i.e., $\hat{\mathbf{y}}_{T+1:T+h} = f(\mathbf{X}_{1:T}, \mathbf{y}_{1:T}|\theta)$, so that the distribution of our output could be as close to the true future values distribution as possible. That is, we wish to find
$\min_\theta \mathbb{E}_\mathbf{X, y} \sum_{t=T+1}^{T+h} \mathrm{KL}\big( y_t || \hat{y}_{t} \big).$
Here, we use Kullback-Leibler (KL) divergence to measure the difference between the distributions of the
true future values $\mathbf{y}_{T+1:T+h}$ and the predictions $\hat{\mathbf{y}}_{T+1:T+h}$. Note that our formulation can be easily extended to multivariate forecasting, in which the output and the corresponding input become multivariate series $\hat{\mathbf{y}}_{T+1:T+h} \in \mathbb{R}^{k \times h}$ and $\mathbf{y}_{1:T} \in \mathbb{R}^{k \times h}$, respectively, where $k$ is the number of forecasting variables,
The related input series is then $\mathbf{X}_{1:T} \in \mathbb{R}^{(n-k) \times T}$, and the overall objective becomes
$\min_\theta \mathbb{E}_{\mathbf{X}_{1:T}, \mathbf{y}_{1:T} } \sum_{t=T+1}^{T+h} \sum_{i=1}^k \mathrm{KL}\big( y_{i, t} || \hat{y}_{i, t} \big).$
In this paper, in order to increase the sample efficiency and maintain a relatively small number of parameters, we will train $d$ separate models to forecast each series individually. 

\subsection{A Distributional Representation of Stocks: Stock2Vec} \label{sec:vec}
In machine learning fields, the categorical variables, if are not ordinal, are often one-hot encoded into a sparse representation. 
i.e., 
$$e: x\mapsto \delta (x, c),$$
where $\delta(x, c)$ is the Kronecker delta, in which each dimension represents a possible category. Let the number of categories of $x$  be $|C|$, then $\delta(x, c)$ is a vector of length $|C|$ with the only element  set to 1 for $x=c$, and all others being zero.
Note that although providing a convenient and simple way of representing categorical variables with numeric values for computation, one-hot encoding has various limitations. 
First of all, it does not place similar categories closer to one another in vector space, within one-hot encoded vectors, all categories are orthogonal to each other thus are totally uncorrelated, i.e., it cannot provide any information on similarity or dissimilarity between the  categories.
In addition, if $|C|$ is large, one-hot encoded vectors can be high-dimensional and often sparse, which means that a prediciton model has to involve  a large number of parameters resulting in inefficient computaitons.
For the cross-sectional data that we use for stock market, 
the number of total interactions between all pairs of stocks increases exponentially with the number of symbols we consider, for example, there are approximately ${500 \choose 2} \approx 0.1$ million pairwise interactions among the S\&P 500 stocks. This number keeps growing exponentially as we add more features to describe the stock price performance
Therefore, trading on cross-sectional signals is remarkably difficult, and approximation methods are often applied. 

We would like to overcome the abovementioned issue by reducing the dimensionality of the categorical variables. Common (linear) dimensionality reduction techniques include the principal component analysis (PCA), singular value decomposition (SVD), which operate by maintaining the first few eigen or singular vectors corresponding to the largest  few engen or singular values
PCA and SVD make efficient use of the statistics from the data and have been proven to be effective in various fields, yet they do not scale well for big matrices (e.g., the computational cost is $\mathcal{O}(n^3)$ for a $n\times n$ matrix), and they cannot adapt to minor changes in the data.
In addition, the unsupervised transformation based on PCA or SVD do not use predictor variable, and hence it is possible that the derived components that serve as surrogate predictors provide no suitable relationship with the target.  
Moreover, since PCA and SVD utilize the first and second moments, they rely heavily on the assumption that the original data have approximate Gaussian distribution, which also limits the effectiveness of their usage.

Neural embedding is another approach to dimensionality reduction. 
Instead of computing and storing global information about the big dataset as in PCA or SVD, neural embedding learning provides us a way to  learn iteratively on a supervised task directly. In this paper, we present a simple probabilistic method, Stock2Vec, 
that learns a dense distributional representation of stocks in a relatively lower dimensional space, and is able to capture the correlations and other more complicated relations between stock prices as well.

The idea is to design such a model whose parameters are the embeddings. We call a mapping $\phi: x \to \mathbb{R}^D$ a $D$-dimensional embedding of $x$, and $\phi(x)$ the embedded representation of $x$. Suppose the transformation is linear, then the embedding representation can be written as $$z = W x = \sum_c w_{c} \delta_{x, c}.$$
The linear embedding mapping is equivalent to an extra fully-connected layer of neural network without nonlinearity on top of the one-hot encoded input. Then each output of the extra linear layer is given as $$z_d = \sum_c w_{c, d} \delta_{x, c} = w_{d}x,$$ where $d$ stands for the index of embedding layer, and $w_{c, d}$ is the weight connecting the one-hot encoding layer to the embedding layer. The number of dimensions $D$ for the embedding layer is a hyperparameter that can be tuned based experimental results, usually bounded between 1 and $|C|$. For our Stock2Vec, as we will introduce in Section \ref{sec:s2v_res}, there are 503 different stocks, and we will map them into a 50-dimensional space.

The assumption of learning a distribuional representation is that the series that have similar or opposite movement tend to correlated with each other, which is consistent with the assumption of CAPM, that the return of a stock is correlated with the market return, which in turn is determined by all stocks' returns in the market. We will learn the embeddings as part of the neural network for the target task of stock prediction. 
In order to 
learn the intrinsic relations among different stocks, we train the deep learning model on data of all symbols over the market, where each datum maintains the features for its particular symbol's own properties, include the symbol itself as a categorical feature,
with the target to predict next day's price. The training objective is to minimize the mean squared error of the predicted prices as usual.

\begin{figure}[htb]
    \centering
    \includegraphics[scale=.8]{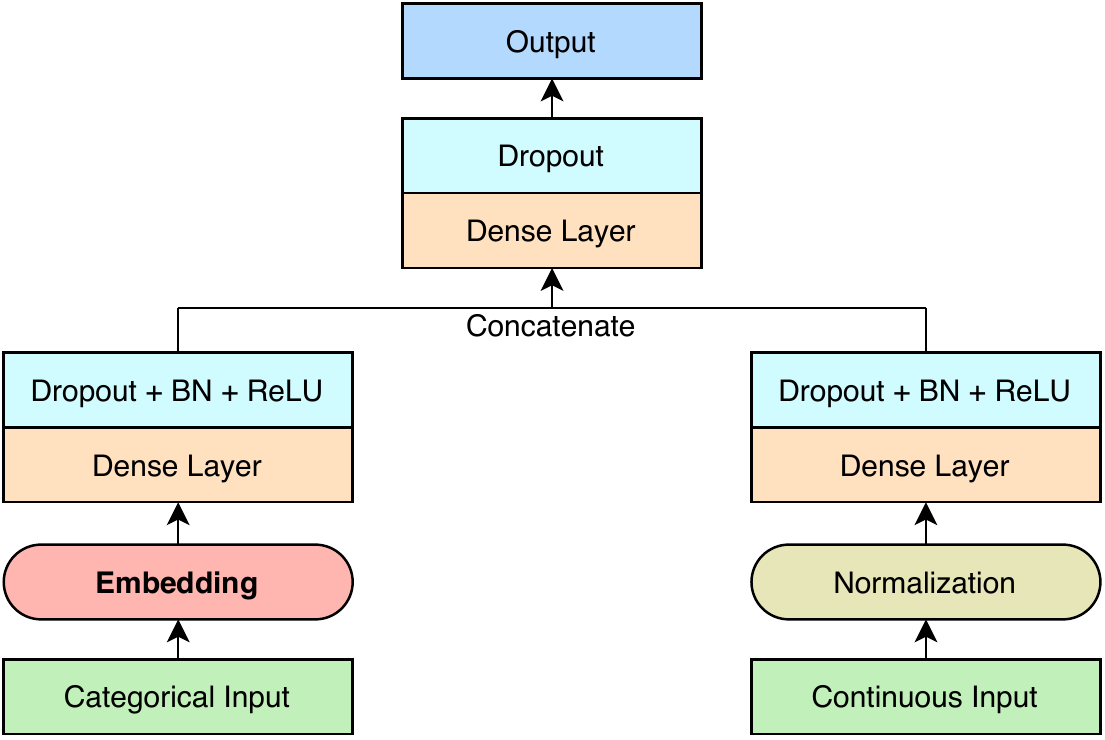}
    \caption{Model Architecture of Stock2Vec.}
    \label{fig:stage1}
\end{figure}

\subsection{Temporal Convolutional Network}


In contrast to standard fully-connected neural networks in which a separate weight describes an interaction between each input and output pair, CNNs share the parameters for multiple mappings.  This is achieved by constructing a collection of kernels (aka filters) with fixed size (which is generally much smaller than that of the input), each consisting of a set of trainable parameters, therefore, the number of parameters is greatly reduced.
Multiple kernels are usually trained and used together, each  specialized in capturing a specific feature from the data.
Note that the so-called convolution operation is technically a cross-correlation in general, which generates linear combinations of a small subset of input, thus focusing on local connectivity.  In CNNs we generally assume that the input data has some grid-like topology, and the same characteristic of the pattern would be the same for every location, i.e., yields the property of equivariance to translation \cite{goodfellow2016deep}.  The size of the output would then not only depend on the size of the input, but also on several settings of the kernels: the stride, padding, and the number of kernels. The stride $s$ denotes the interval size between two consecutive convolution centers, and can be thought of as downsampling the output. Whereas with padding, we add values (zeros are used most often) at the boundary of the input, which is primarily used to control the output size, but as we will show later, it can also be applied to manage the starting position of the convolution operation on the input. The number of kernels adds another dimensionality on the output, and is often denoted as the number of channels.

\subsubsection{1D Convolutional Networks}
Sequential data often display long-term correlations and can be though of as a 1D grid with  samples taken at regular time intervals. CNNs have shown success in time series applications, in which the 1D convolution is simply an operation of sliding dot products between the input vector and the kernel vector. However, we  make several modifications to traditional  1D convolutions according to recent advances. 
The detailed building blocks of our temporal CNN components are illustrated in the following  subsections. 

\subsubsection{Causal Convolutions} \label{sec:causal}
\begin{figure}[htb]
    \centering
    \begin{subfigure}[b]{0.45\textwidth}
        \includegraphics[width=\textwidth]{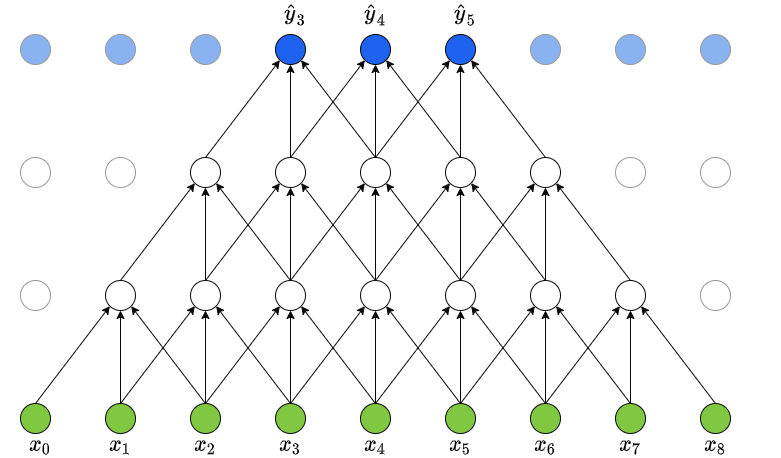}
        \caption{standard (non-causal)}
        \label{fig:causal}
    \end{subfigure}
    \hfill
    \begin{subfigure}[b]{0.45\textwidth}
        \includegraphics[width=\textwidth]{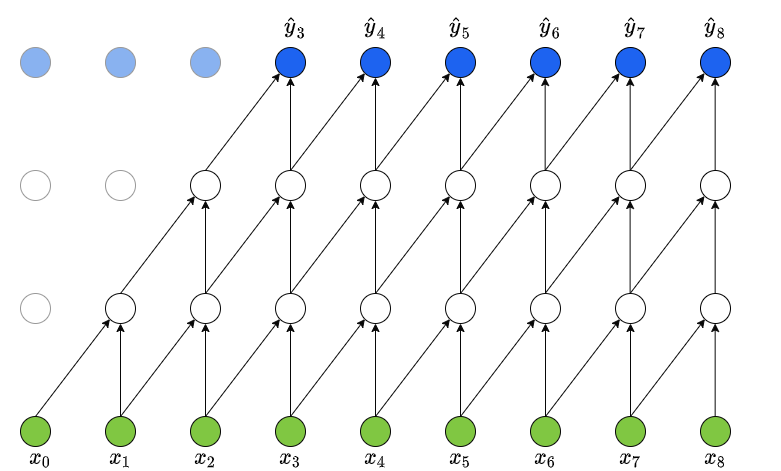}
        \caption{causal}
    \end{subfigure}
    \caption{Visualization of a stack of 1D convolutional layers, non-causal v.s. causal.}
    \label{fig:non_vs_causal}
\end{figure}

As we mentioned above, in a traditional 1D convolutional layer, the filters are slided across the input series. As a result, the output is related to the connection structure between the inputs before and after it. As shown in Figure \ref{fig:non_vs_causal}(a), by applying a filter of width 2 without padding, the predicted outputs $\hat{x}_1, \cdots, \hat{x}_T$ are generated using the input series $x_1, \cdots, x_T$. The most severe problem within this structure is that we use the future to predict the past, e.g., we have used $x_2$ to generate $\hat{x}_1$, which is not appropriate in time series analysis. To avoid the issue, causal convolutions are used, in which the output $x_t$ is convoluted only with input data which are earlier and up to time $t$ from the previous layer. We achieve this by explicitly zero padding of length $(kernel\_size - 1)$ at the beginning of input series, as a result, we actually have shifted the outputs for a number of time steps. In this way, the prediction at time $t$ is only allowed to connect to historical information, i.e., in a causal structure, 
thus we have prohibited the future affecting the past and avoided information leakage. 
The resulted causal convolutions is visualized in Figure \ref{fig:non_vs_causal}(b).

\subsubsection{Dilated Convolutions} \label{sec:dilated}
\begin{figure}[htb]
    \centering
    \begin{subfigure}[b]{0.85\textwidth}
        \includegraphics[width=\textwidth]{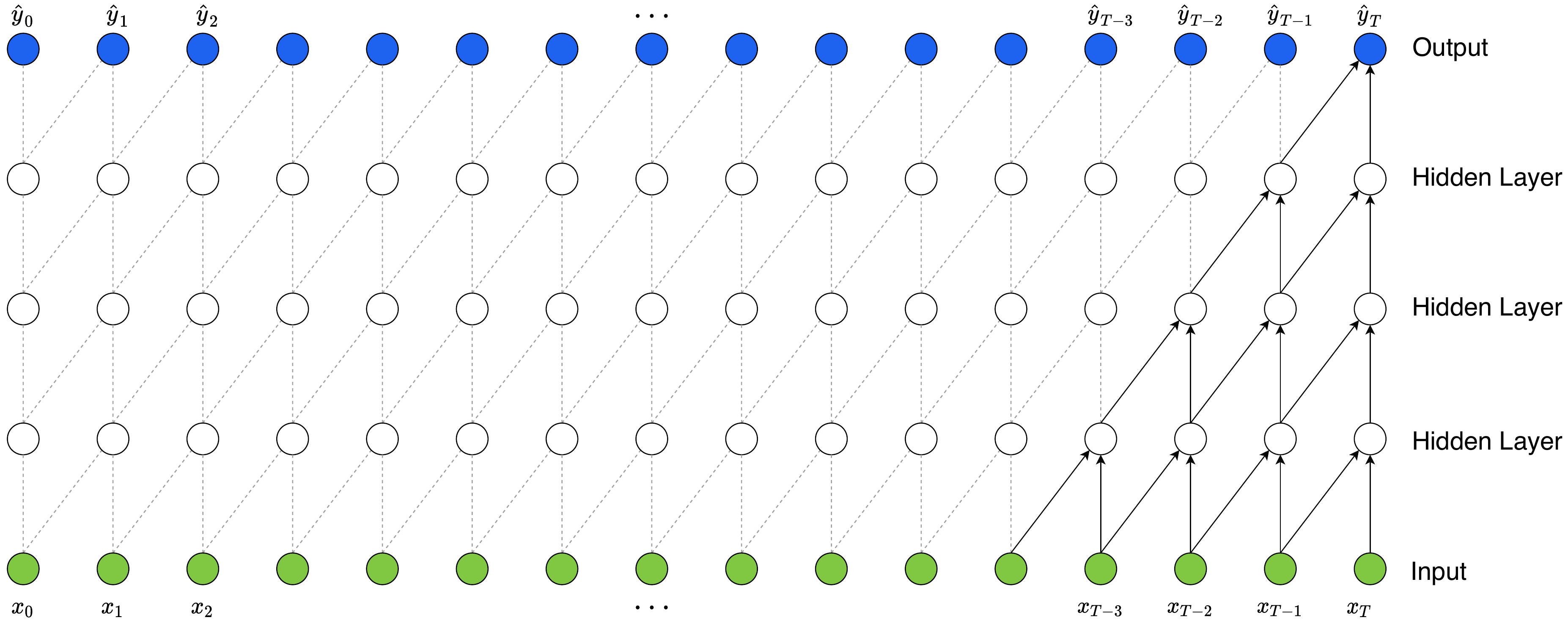}
        \caption{Non-dilated}
        \label{fig:nondilated}
    \end{subfigure}
    \par\bigskip
    \begin{subfigure}[b]{0.85\textwidth}
        \includegraphics[width=\textwidth]{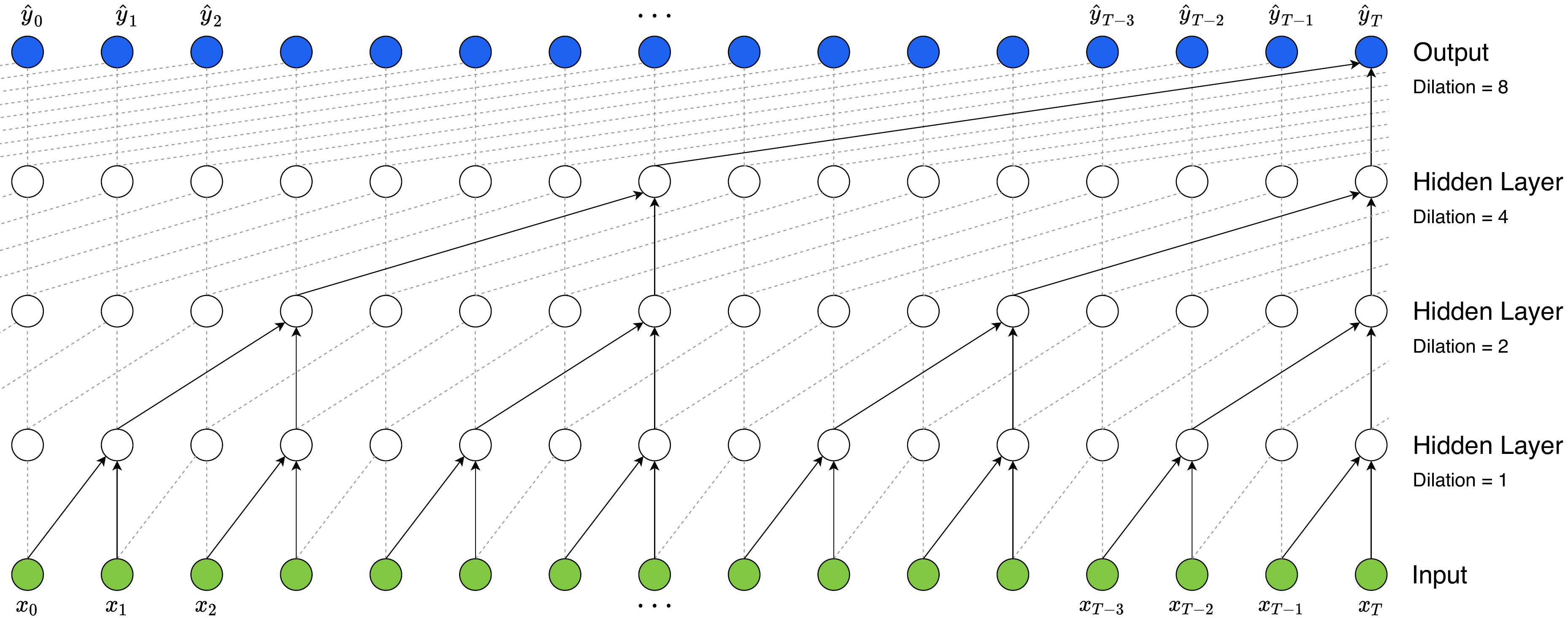}
        \caption{Dilated}
        \label{fig:dilated}
    \end{subfigure}
    \caption{Visualization of a stack of causal convolutional layers, non-dilated v.s. dilated.}
    \label{fig:non_vs_dilated}
\end{figure}

Time series often exhibits long-term autoregressive dependencies. With neural network models hence, we require for the receptive field of the output neuron to be large. That is, the output neuron should be connected with the neurons that receive the input data from many time steps in the past.  
A major disadvantage of the aforementioned basic causal convolution is that in order to have large receptive field,  either very large sized filters are required, or those need to be stacked in many layers. With the former, the merit of CNN architecture is  lost, and with the latter, the model can become computationally intractable.
Following \cite{oord2016wavenet}, we adopted the dilated convolutions in our model instead, which is defined as $$F(s) = (\mathbf{x}\ast_d f)(s) = \sum_{i=0}^{k-1}f(i)\cdot \mathbf{x}_{s-d\times i}, $$ where $x\in \mathbb{R}^T$ is a 1-D series input, and $f:\{ 0, \cdots, k-1\} \to \mathbb{N}$ is a filter of size $k$, $d$ is called the dilation rate, and $(s-d\times i)$ accounts for the direction of the past. In a dilated convolutional layer, filters are not convoluted with the inputs in a simple sequential manner, but instead skipping a fixed number ($d$) of inputs in between. By increasing the dilation rate multiplicatively as the layer depth (e.g., a common choice is  $d=2^j$ at depth $j$), we increase the receptive field 
exponentially, i.e., there are $2^{l-1}k$ input in the first layer that can affect the output in the $l$-th hidden layer.  Figure \ref{fig:non_vs_dilated} compares non-dilated and dilated causal convolutional layers.

\subsubsection{Residual Connections} \label{sec:resnet}
\begin{figure}[htb]
    \centering
    \begin{subfigure}[b]{0.475\textwidth}
        \hspace{.75in}
        \includegraphics[scale=.8]{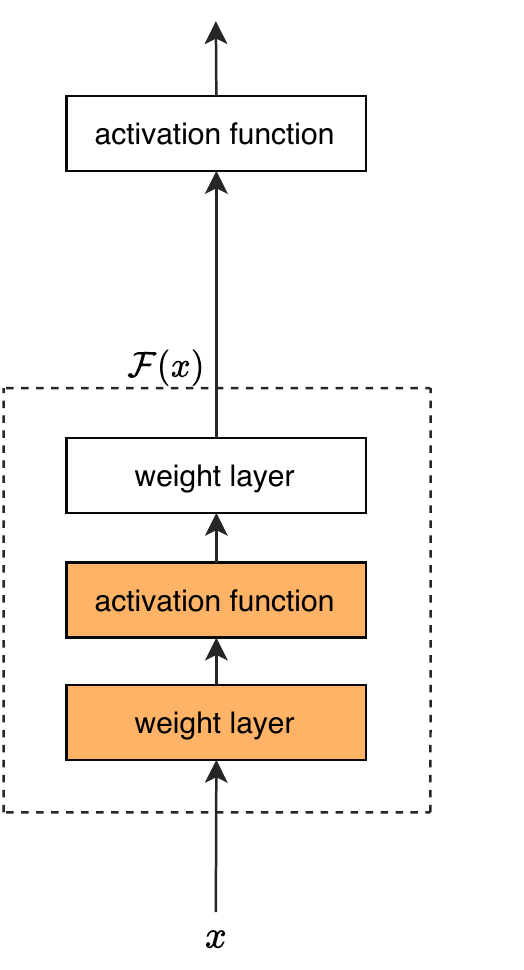}
        \caption{A standard block}
    \end{subfigure}
    ~
    \begin{subfigure}[b]{0.475\textwidth}
        \hspace{.75in}
        \includegraphics[scale=.8]{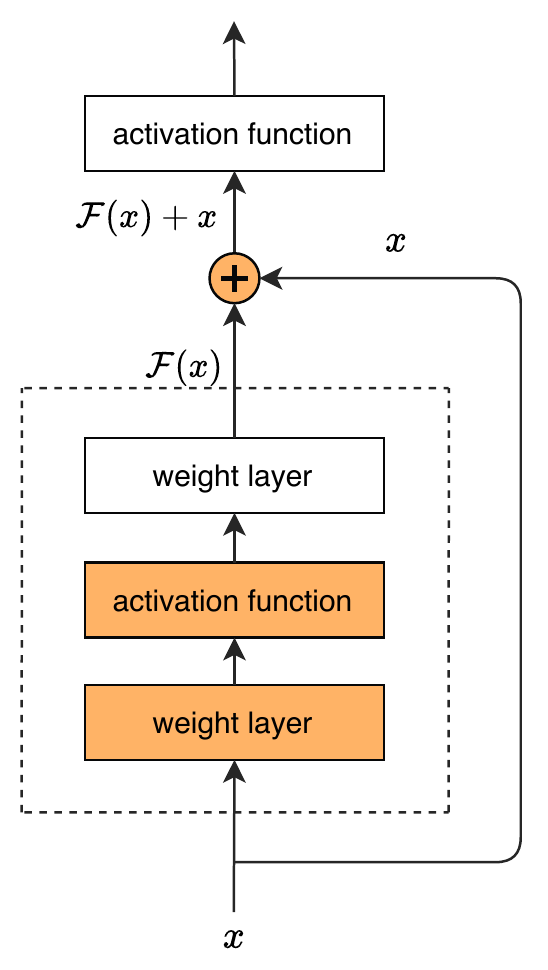}
        \caption{A residual block}
    \end{subfigure}
    \caption{Comparison between a regular block and a residual block. In the latter, the convolution is short-circuited.}
    \label{fig:resblock}
\end{figure}

In traditional neural networks, each layer feeds into the next. In a network with residual blocks, by utilizing skip connections, a layer may also short-cut to jump over several others. The use of residual network (ResNet) \cite{he2016deep} has been proven to be very successful and become the standard way of building deep CNNs. 
The core idea of ResNet is the usage of shortcut connection which skips one or more layers and directly connects to later layers (which is the so-called identity mapping), in addition to the standard layer stacking connection $\mathcal{F}$. 
Figure \ref{fig:resblock} illustrates a residual block, which is the basic unit in ResNet. A residual block consists of the abovementioned two branches, and its output is then $g(\mathcal{F}(x)+x)$, where $x$ denotes the input to the residual block, and $g$ is the activation function.

By reusing activation from a previous layer until the adjacent layer learns its weights, CNNs can effectively avoid the problem of vanishing gradients. In our model, we implemented  double-layer skips.

\subsection{The Hybrid Model} \label{sec:hybrid_intro}

\begin{figure}[htb]
    \centering
    \includegraphics[width=.8\textwidth]{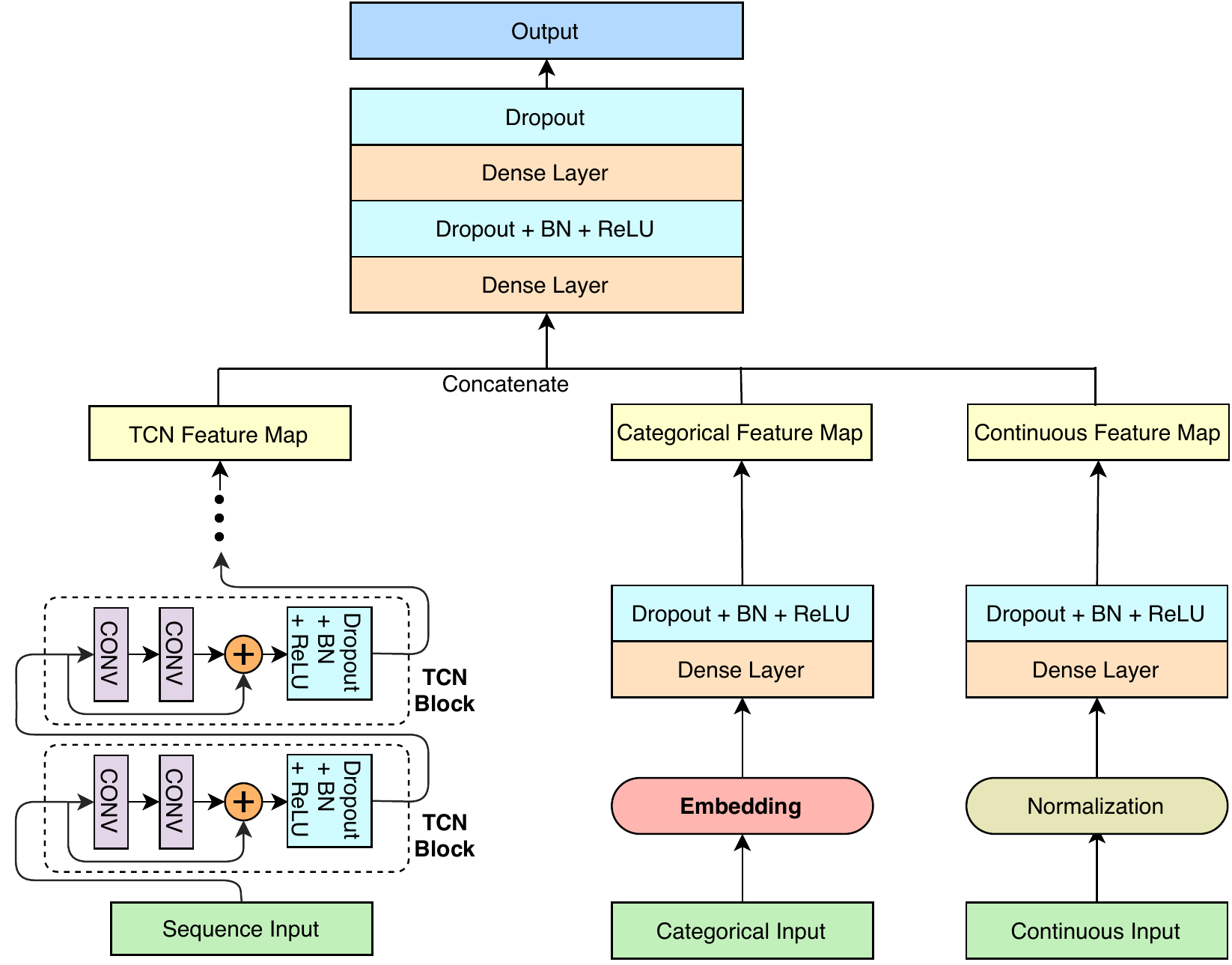}
    \caption{The full model architecture of hybrid TCN-Stock2Vec.}
    \label{fig:hybrid2}
\end{figure}

Our overall prediction model is constructed as a hybrid, combining Stock2Vec embedding approach with an advanced implementation of TCN, schematically represented on  Figure \ref{fig:hybrid2}.
Compared with Figure \ref{fig:stage1}, it contains an additional TCN module. 
However, instead of producing the final prediction outputs of size 1, we  let the TCN module output a vector as a feature map that contains information extracted from the temporal series. As a result, it adds a new source of features, which can be We concatenated with the learned Stock2Vec features. 
Note that the TCN module can be replaced by any other architecture that learns temporal patterns, for example, LSTM-type network. 
Finally, a series of fully-connected layers (referred to as ``head layers'') are applied to the combined features producing the final prediction output. Implementation details are discussed in Section \ref{sec:s2v_hyper}.

Note that in each TCN block, the convolutional layers use dropout in order to limit the influence that earlier data have on learning \cite{srivastava2014dropout, gal2016dropout}. It is then followed by a batch normalization layer \cite{ioffe2015batchnorm}. Both dropout and batch normalization provide  a regularization effect that avoids overfitting. The most widely used activation function, the rectified linear unit (ReLU) \cite{nair2010rectified} is used after each layer except for the last one.

%% file: stock/data.tex
The case study is based on daily trading data for 503 assets listed on S\&P 500 index, downloaded from Yahoo!finance for the period of 2015/01/01--2020/02/18 (out of 505 assets listed on \url{https://en.wikipedia.org/wiki/List_of_S\%26P_500_companies}, two did not have data spanning the whole period).
%
Following the literature, we use the next day's
closing price as the target label for each asset, while the adjusted closing prices up until the current date can be used as inputs.
In addition, we  also use as augmented features the downloaded open/high/low prices and volume data for calculating some commonly used technical indicators that reflect price variation over time. 
In our study, eight commonly used technical indicators \cite{murphy1999technical} are selected, which are described in Table \ref{tab:techind}. 
As we discussed in Section \ref{sec:s2v_intro}, 
it has also been shown in the literature that  assets' media exposure and the corresponding text sentiment are highly correlated with the stock prices. 
To account for this, we acquired another set of features through the Quandl API.
The database ``FinSentS Web News Sentiment'' (\url{https://www.quandl.com/databases/NS1/}) is used in this study. The queried dataset includes the daily number of news articles about each stock, as well as the sentiment score that measures the texts used in media, based on proprietary algorithms for web scraping and natural language processing.

We further extracted several date/time related variables for each entry to explicitly capture the seasonality, these features include month of year, day of month, day of week, etc. All of the above-mentioned features
are all dynamic features that are time-dependent. In addition, we gathered a set of static features that are time-independent. Static covariates (e.g., the symbol name, sector and industry category, etc.) could assist the feature-based learner to capture series-specific information such as the scale level and trend for each series. 
The distinction between dynamic and static features is important for model architecture design, since it is unnecessary to process the static covariates by RNN cells or CNN convolution operations for capturing temporal relations (e.g., autocorrelation, trend, and seasonality, etc.).

\begin{table*}[ht]
\centering
\resizebox{\textwidth}{!}{%
\begin{tabular}{lccl|}
\hline
\textbf{Technical Indicators} & \textbf{Category} & \textbf{Description} \\ \hline \hline
Moving average convergence or divergence (MACD) & Trend &  Reveals price change in strength, direction and trend duration \\  
Parabolic Stop And Reverse (PSAR) & Trend &  Indicates whether the current trend is to continue or to reverse \\  
Bollinger Bands (BB\textsuperscript{\textregistered}
) & Volatility & Forms a range of prices for trading decisions \\  
Stochastic Oscillator (SO) & Momentum & Indicates turning points by comparing the price to its range \\   
Rate Of Change (ROC) & Momentum & Measures the percent change of the prices \\  
On-Balance Volume (OBV) & Volume &  Accumulates volume on price direction to confirm price moves \\  
Force Index (FI) & Volume & 
Measures the amount of strength behind price move \\ \hline
\end{tabular}%
}
    \caption{Description of technical indicators used in this study.}
    \label{tab:techind}
\end{table*}

Note that the features can also be split into categorical and continuous.
Each of the categorical features is mapped to dense numeric vectors via embedding, in particular, the vectors embedded from the stock name as a categorical feature are called Stock2Vec. 
We scale all continuous features (as well as next day's price as the target) to between 0 and 1, since it is widely accepted that neural networks are hard to train and are sensitive to input scale \cite{glorot2010understanding, ioffe2015batchnorm}, while some alternative approaches, e.g., decision trees, are scale-invariant  \cite{covington2016deep}.
It is important to note that we performed scaling separately on each asset, i.e., linear transformation is performed so that the lowest and highest price for asset A  over the training period is 0 and 1 respectively. Also note scaling statistics are obtained with the training set only, which prevents  leakage of information from the test set, avoiding introduction of  look-ahead bias.

\begin{figure}[htb]
    \centering
    \includegraphics[width=.7\textwidth]{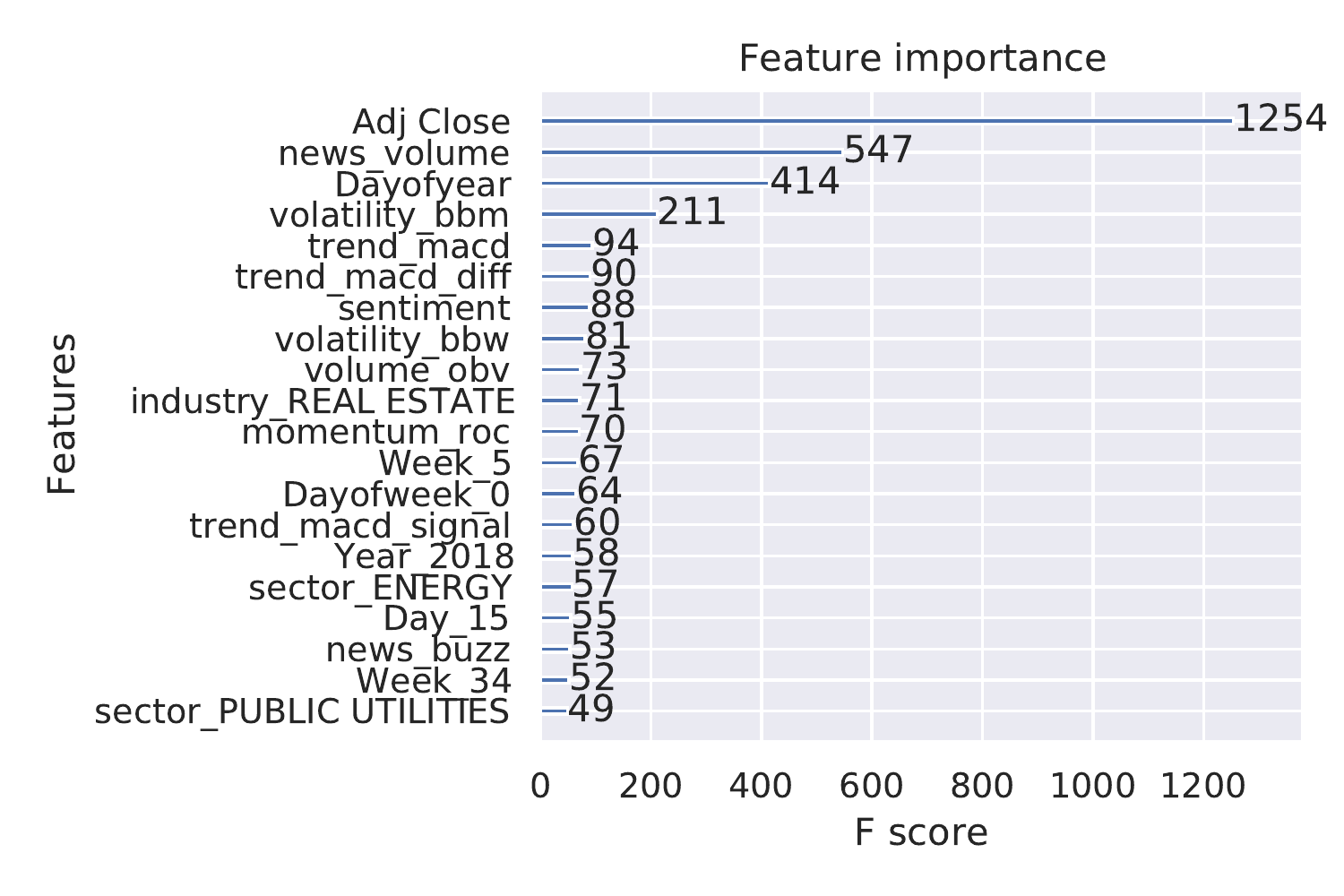}
    \caption{Feature importance plot of XGBoost model.}
    \label{fig:feature_importance}
\end{figure}

As a tentative illustration, Figure \ref{fig:feature_importance} shows the most important 20 features for predicting next day's stock price, according to the XGBoost model we trained for benchmarking.

In our experiments, the data are split into training, validation and test sets. The last 126 trading days of data are used as the test set, cover the period from 2019/08/16 to 2020/02/18, and include 61000 samples. The rest data are used for training the model, in which the last 126 trading days, from 2019/02/15 to 2019/08/15, are used as validation set, while the first 499336 samples, cover the period from 2015/01/02 to 2019/02/14, form the training set. Table \ref{tab:datsummary} provides a summary of the datasets we used in this research.

\begin{table*}[ht]
\centering
    \caption{Dataset summary.}
\footnotesize
\begin{tabular}{lccc}
\hline
& Training set & Validation set & Test set \\ \hline \hline
Starting date & 2015/01/02 & 2019/02/15 & 2019/08/16 \\
End date & 2019/02/14 & 2019/08/15 & 2020/02/18 \\
Sample size & 499336 & 61075 & 61000 \\ \hline
\end{tabular}%
    \label{tab:datsummary}
\end{table*}

%% file: stock/results.tex
\subsection{Benchmark Models, Hyperparameters and Optimization Strategy} \label{sec:s2v_hyper}

In the computational experiments below we compare performance of seven models.. Two models are based on time series analysis only (TS-TCN and TS-LSTM), two use static feature only (random forest \cite{breiman2001random} and XGBoost \cite{chen2016xgboost}), pure Stock2Vec model and finally, two versions of the proposed hybrid model (LSTM-Stock2Vec and TCN-Stock2Vec). This way we can evaluate the effect of different model architectures and data features. Specifically, we are interested in evaluating whether employing feature embedding leads to improvement (Stock2Vec vs random forest and XGBoost) and whether a further improvement can be achieved by incorporating time-series data in the hybrid models.


Random forest and XGBoost are ensemble models that deploy enhanced bagging and gradient boosting, respectively. 
We pick these two models since both have shown powerful predicting ability and achieved  state-of-the-art performance in various fields. Both  are tree-based models that are invariant to scales and perform split on one-hot encoded categorical inputs, which is suitable for comparison with embeddings in our Stock2Vec models. 
We built 100 bagging/boosting trees for these two models. 
LSTM and TCN models are constructed based on  pure time series data, i.e., the inputs and outputs are single series, without any other feature as augmented series. In later context, we call these two models TS-LSTM and TS-TCN, respectively. 
The Stock2Vec model is a fully-connected neural network with embedding layers for all categorical features, it has the exactly same inputs as XGBoost and random forest. As we introduced in Section \ref{sec:hybrid_intro}, our hybrid model combines the Stock2Vec model with an extra TCN module to learn the temporal effects. And for comparison purpose, we also evaluated the hybrid model with LSTM as the temporal module. We call them TCN-Stock2Vec and LSTM-Stock2Vec correspondingly. 


Our deep learning models are implemented in PyTorch \cite{paszke2017pytorch}. In Stock2Vec, the embedding sizes are set to be half of the original number of categories, thresholded by 50 (i.e., the maximum dimension of embedding output is 50). These are just heuristics as there is no common standard for choosing the embedding sizes. We concatenate the continuous input with the outputs from embedding layers, followed by two layers of fully-connected layers, with sizes of 1024 and 512, respectively. The dropout rates are set to 0.001 and 0.01 for the two hidden layers correspondingly.

For the RNN module, we implement two-layer stacked LSTM, i.e., in each LSTM cell (that denotes a single time step), there are two LSTM layers sequentially connected, and each layer consists of 50 hidden units. We need an extra fully-connected layer to control the output size for the temporal module, depending on whether to obtain the final prediction as in TS-LSTM (with output size to be 1), or a temporal feature map as in LSTM-Stock2Vec. We set the size of temporal feature map to be 30 in order to compress the information for both LSTM-Stock2Vec and TCN-Stock2Vec.
In TCN, we  use another convolutional layer to achieve the same effect.
To implement the TCN module, we build a 16-layer dilated causal CNN as the component that focuses on capturing the autoregressive temporal relations from the series own history. Each layer contains 16 filters, and each filter has a width of 2. Every two consecutive convolutional layers form a residual block after which the previous inputs are added to the flow.
The dilation rate increases exponentially along every stacked residual blocks, i.e., to be $1, 2, 4, 8, \cdots, 128$, which allows our TCN component to capture the autoregressive relation for more than half a year (there are 252 trading days in a year). Again, dropout (with probability $0.01$), batch normalization layer and ReLU activation are used for each TCN block. 

The MSE loss is used for all models. The deep learning models were trained using stochastic gradient descent (SGD), with batch size of 128. In particular, the Adam optimizer \cite{kingma2014adam} with initial learning rate of $10^{-4}$ was used to train TS-TCN and TS-LSTM. 
To train Stock2Vec, we deployed the super-convergence scheme as in  \cite{smith2019super} and used cyclical learning rate over every 3 epochs, with a maximum value of $10^{-3}$. 
In the  two hybrid models, while the weights of the head layers were randomly initialized as usual, we loaded the weights from pre-trained Stock2Vec and TS-TCN/TS-LSTM for the corresponding modules. By doing this, we have applied transfer learning scheme \cite{pan2009survey, bengio2012deep, long2017deep} and wish the transferred modules have the ability to effectively process features from the beginning. 
The head layers were trained for 2 cycles (each contains 2 epochs) with maximum learning rate of $3\times 10^{-4}$ while the transferred modules were frozen. After this convergence, the entire network was fine-tuned for 10 epochs by standard Adam optimizer with learning rate of  $10^{-5}$, during which an early stopping paradigm \cite{yao2007early} was applied to retrieve the model with smallest validation error. 
We select the hyperparemeters based upon the model performance on the validation set. 

\subsection{Performance Evaluation Metrics}
To evaluate the performance of our forecasting model, three commonly used evaluation criteria are used in this study: (a) the root mean square error (RMSE), (b) the mean absolute error (MAE), (c) the mean absolute percentage error (MAPE), (d) the root mean square percentage error (RMSPE):
\begin{align}
    \textrm{RMSE} & = \sqrt{\frac{1}{H}\sum_{t=1}^H (y_t - \hat{y}_t)^2} \\
    \textrm{MAE} & = \frac{1}{H}\sum_{t=1}^H \big|y_t - \hat{y}_t \big| \\
    \textrm{MAPE} &= \frac{1}{H}\sum_{t=1}^H \Big|\frac{y_t - \hat{y}_t}{y_t} \Big| \times 100 \\
    \textrm{RMSPE} &= \sqrt{ \frac{1}{H}\sum_{t=1}^H \Big|\frac{y_t - \hat{y}_t}{y_t} \Big|^2 } \times 100
\end{align}
where $y_t$ is the actual target value for the $t$-th observation, $\hat{y}_t$ is the predicted value for the corresponding target, and $H$ is the forecast horizon. 

The RMSE is the most popular measure for the error rate of regression models, as $n\to \infty$, it converges to the standard deviation of the theoretical prediction error. However, the quadratic error may not be an appropriate evaluation criterion for all prediction problems, especially in the presence of large outliers. In addition, the RMSE depends on scales, and is also sensitive to outliers. The MAE considers the absolute deviation as the loss and is a more ``robust'' measure for prediction, since the absolute error is more sensitive to small deviations and much less sensitive to large ones than the squared error. However, since the training process for many learning models are based on squared loss function, the MAE could be (logically) inconsistent to the model optimization selection criteria. The MAE is also scale-dependent, thus not suitable to compare prediction accuracy across different variables or time ranges. In order to achieve scale independence, the MAPE measures the error proportional to the target value, while instead of using absolute values, the RMSPE can be seen as the root mean squared version of MAPE. The MAPE and RMSPE however, are extremely unstable when the actual value is small (consider the case when the denominator  or close to 0). We will consider all four measures mentioned here to have a more complete view of the performance of the models considering the limitations of each performance measure. In addtion, we will compare the running time as an additional evaluation criterion. 



\subsection{Stock2Vec: Analysis of Embeddings}
As we introduced in Section \ref{sec:s2v_method}, 
the main goal of training Stock2Vec model is to learn the intrinsic relationships among stocks, where similar stocks are close to each other in the embedding space, so that we can deploy the interactions from cross-sectional data, or more specifically, the market information, to make better predictions. To show this is the case, we extract the weights of the embedding layers from the trained Stock2Vec model, map the weights down to two-dimensional space with a manifold by using PCA, and visualize the entities to look at how the embedding spaces look like.
Note that besides Stock2Vec, we also learned embeddings for other categorical features. 

\begin{figure}[htb]
    \centering
    \begin{subfigure}[b]{0.485\textwidth}
        \includegraphics[width=\textwidth]{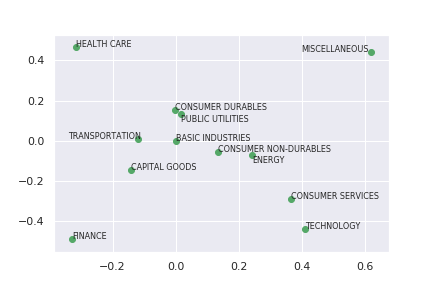}
        \caption{Visualization of learned embeddings for sectors, projected to 2-D spaces using PCA.}
    \end{subfigure}
    ~
    \begin{subfigure}[b]{0.485\textwidth}
        \includegraphics[width=\textwidth]{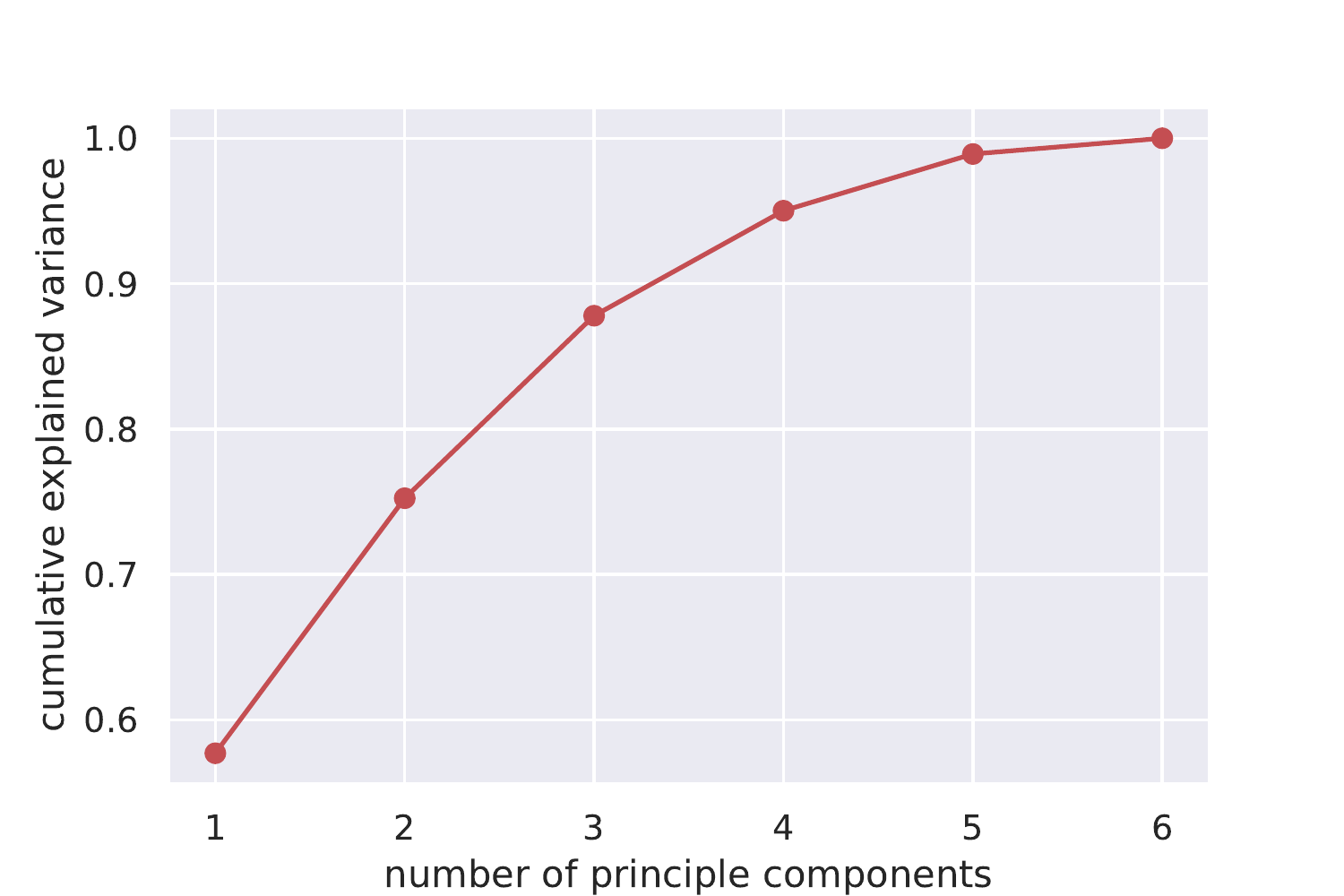}
        \caption{The cumulative explained variance ratio for each of the principal components, sorted by the singular values}
    \end{subfigure}
    \caption{PCA on the learned embeddings for Sectors}
    \label{fig:pca_sector}
\end{figure}

Figure \ref{fig:pca_sector}(a) shows the first two principal components of the sectors.
Note that here the first two components account for close to 75\% of variance. We can generally observe that \emph{Health Care, Technology/Consumer Services} and \emph{Finance} occupy the opposite corners of the plot, i.e., represent unique sectors most dissimilar from one another. On the other hand a collection of more traditional sectors: \emph{Public Utilities, Energy, Consumer Durables and Non-Durables, Basic Industries} generally are grouped closer together. The plot, then, allows for a natural interpretation  which is in accordance with our intuition, indicating that the learned embedding can be expected to be reasonable.


\begin{figure}[htb]
    \centering
    \begin{subfigure}[b]{0.485\textwidth}
        \includegraphics[width=\textwidth]{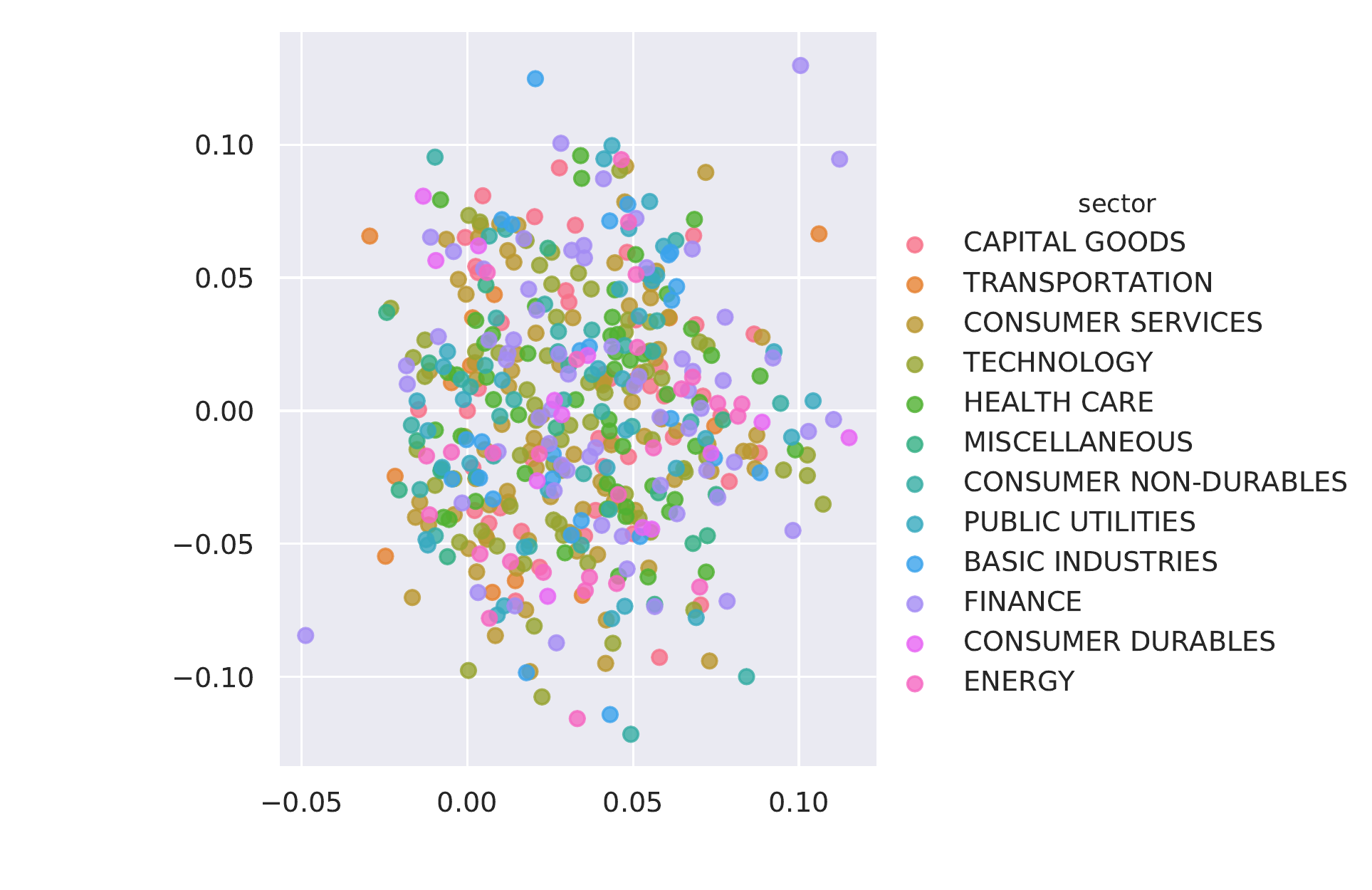}
        \caption{Visualization of Stock2Vec (colored by sectors), projected to 2-D spaces using PCA.}
    \end{subfigure}
    ~
    \begin{subfigure}[b]{0.485\textwidth}
        \includegraphics[width=\textwidth]{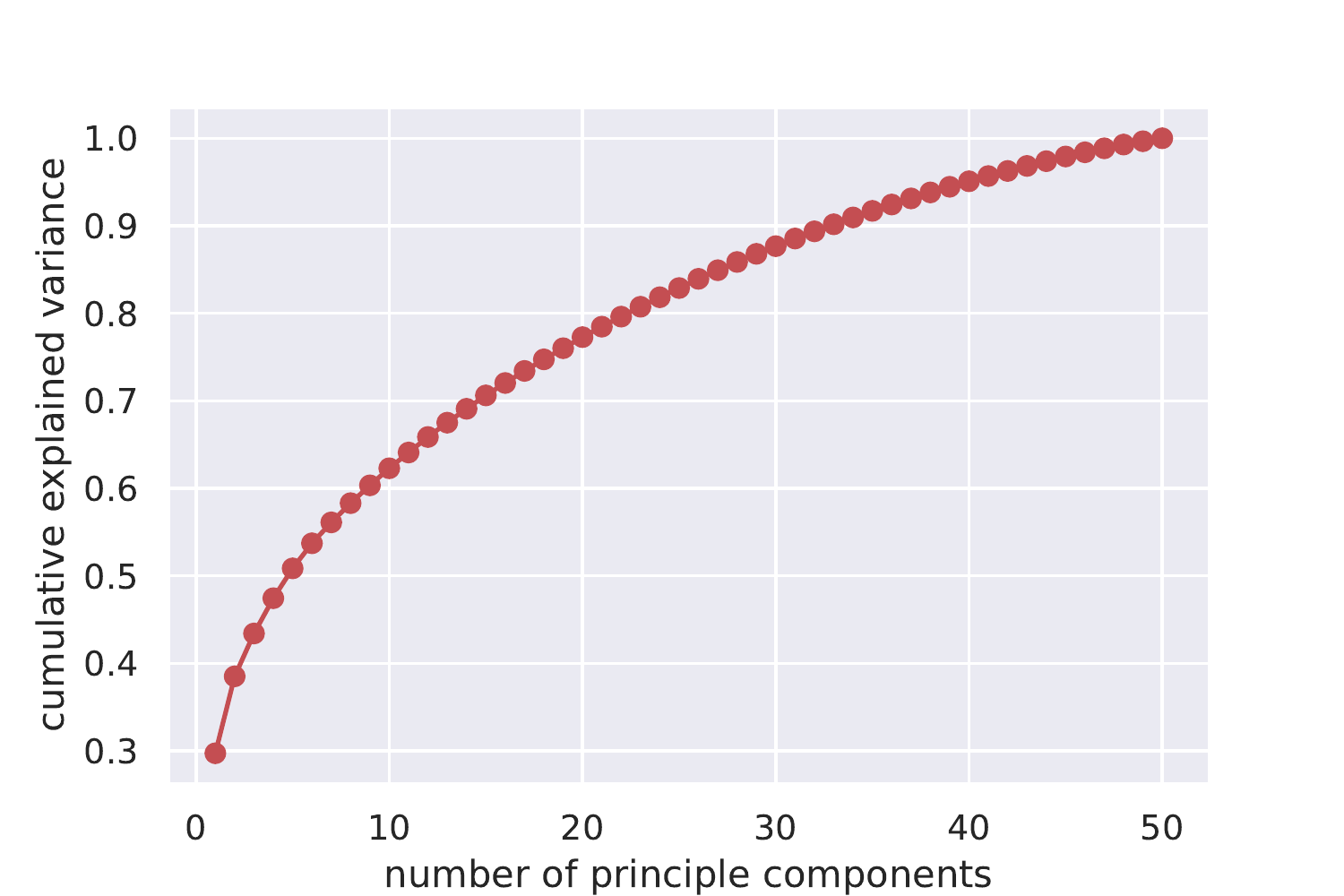}
        \caption{The cumulative explained variance ratio for each of the principal components, sorted by the singular values}
    \end{subfigure}
    \caption{PCA on the learned Stock2Vec embeddings}
    \label{fig:s2v}
\end{figure}

Similarly, from the trained Stock2Vec embeddings, we can obtain a 50-dimensional vector for each separate stock. We simialrly visualize the learned Stock2Vec with PCA in Figure \ref{fig:s2v}(a), and color each stock by the sector it belongs to. 
It is important to note that in this case, the first two components of PCA only account for less than 40\% of variance. In other words, in this case, the plotted groupings do not represent the learned information as well as in the previous case. Indeed, when viewed all together, individual assets do not exhibit readily discernible patterns. This is not necessarily an indicator of deficiency of the learned embedding, and instead suggests that two dimensions are not sufficient in this case.

\begin{figure}[htb]
    \centering
    \begin{subfigure}[b]{0.485\textwidth}
        \includegraphics[width=\textwidth]{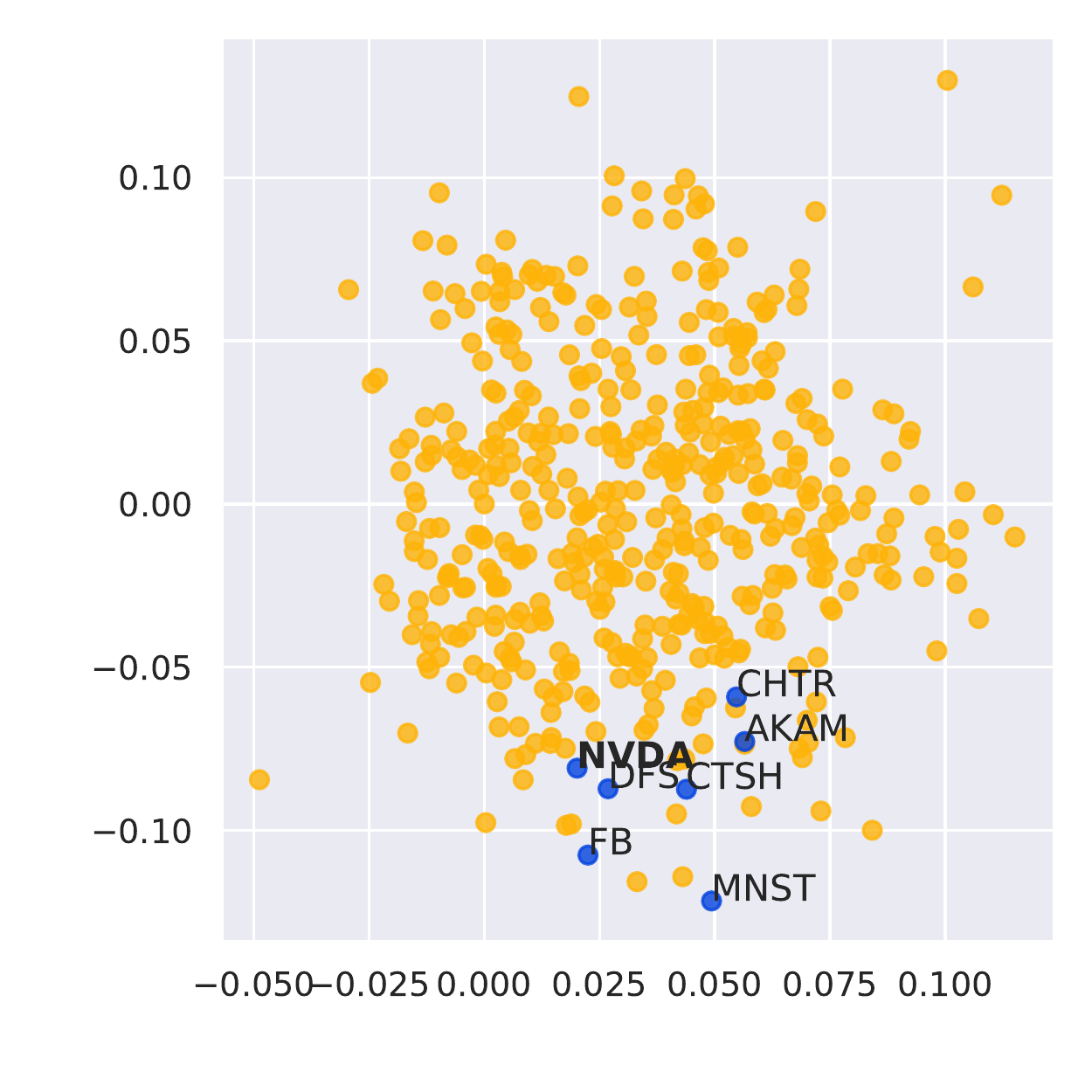}
        \caption{Nearest neighbors of NVDA, which are: 1). 'MNST', Monster, fast growing; 2) 'FB', Facebook, IT; 3)'DFS', Discover Bank, fast growing; 4) 'AKAM', Akamai, IT; 5) 'CTSH', Cognizant Tech Solutions, IT; 6) 'CHTR', Charter, communication services. }
        \label{fig:nvda}
    \end{subfigure}
    ~
    \begin{subfigure}[b]{0.485\textwidth}
        \includegraphics[width=\textwidth]{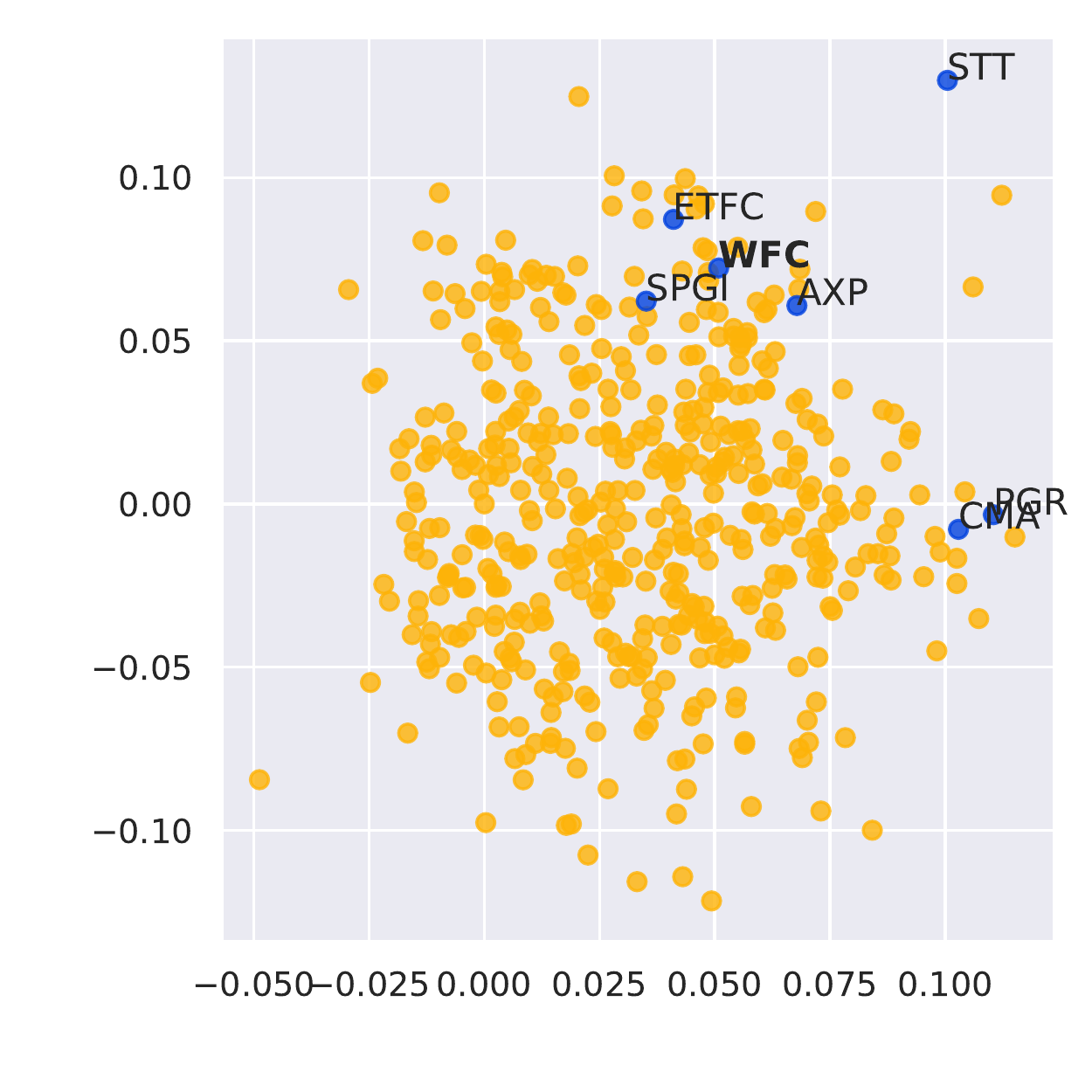}
        \caption{Nearest neighbors of WFC, which are: 1) 'ETFC': E-Trader, financial; 2) 'STT': State Street Corp., bank; 3) 'CMA': Comerica, bank; 4) 'AXP': Amex, financial; 5) 'PGR': Progressive Insurance, financial; 6) 'SPGI', S\&P Global, Inc., financial \& data}
        \label{fig:wfc}
    \end{subfigure}
    \caption{Nearest neighbors of Stock2Vec based on similarity between stocks.}
    \label{fig:neighbors}
\end{figure}

However, lots of useful insight can be gained from the distributed representations, for instance, we could consider the similarities between stocks in the learned vector space is an example of these benefits as we will show below, 

To reveal some additional  insights from the similarity distance, we sort the pairwise cosine distance (in the embedded space) between the stocks in the ascending order. 
In Figure \ref{fig:nvda}, we plot the ticker ``NVDA'' (Nvidia) as well as its six nearest neighbors in the embedding space. The six companies that are closest to Nvidia, according to the embeddings of learned weights, are either of the same type (technology companies) with Nvidia: Facebook, Akamai, Cognizant Tech Solutions, Charte; or fast growing during the past ten years (was the case for Nvidia during the tested period):  Monster, Discover Bank.  
Similarly, we plot the ticker of Wells Fargo (``WFC'') and its 6 nearest neighbors in Figure \ref{fig:wfc}, all of which are either banks or companies that provide other financial services. 

These observations suggest are another indicator that Stock2Vec can be expected to learn some useful information, and indeed is capable of coupling together insights from a number of unrelated sources, in this case, asset sector and it's performance.   
%
%
%

%
The following points must be noted here.   First, most of the nearest neighbors are not the closest points in the two-dimensional plots due to the imprecision of mapping into two-dimensions.
%
Secondly, although the nearest neighbors are meaningful for many companies as the results either are in the same sector (or industry), or present similar stock price trend in the last a few years, this insight does not hold true for all companies, or the interpretation can be hard to discern. For example, the nearest neighbors of Amazon.com (AMZN) include transportation and energy companies (perhaps due to its heavy reliance on these industries for its operation) as well as technology companies. 
%
Finally, note that there exist many other visualization techniques for projection of high dimensional vectors onto 2D spaces that could be used here instead of PCA, for example,  t-SNE \cite{maaten2008visualizing} or UMAP \cite{mcinnes2018umap}. However, neither provided visual improvement of the grouping effect over Figure \ref{fig:s2v}(a) and hence we do not present those results here.


Based on the above observations, Stock2Vec provides several benefits: 
1) reducing the dimensionality of categorical feature space, thus the computational performance is improved with smaller number of parameter,
2) mapping the sparse high-dimensional one-hot encoded vectors onto dense distributional vector space (with lower dimensionality), as a result, similar categories are learned to be placed closer to one another in the embedding space, unlike in one-hot encoding vector space where every pairs of categories yield the same distance and are orthogonal to each other. Therefore, the outputs of the embedding layers could be served as more meaningful features, for later layers of neural networks to achieve more effective learning. Not only that, the meaningful embeddings can be used for visualization, provides us more interpretability of the deep learning models. 

\subsection{Prediction Results}

\begin{figure}[tbh!]
    \centering
    \includegraphics[width=0.7\textwidth]{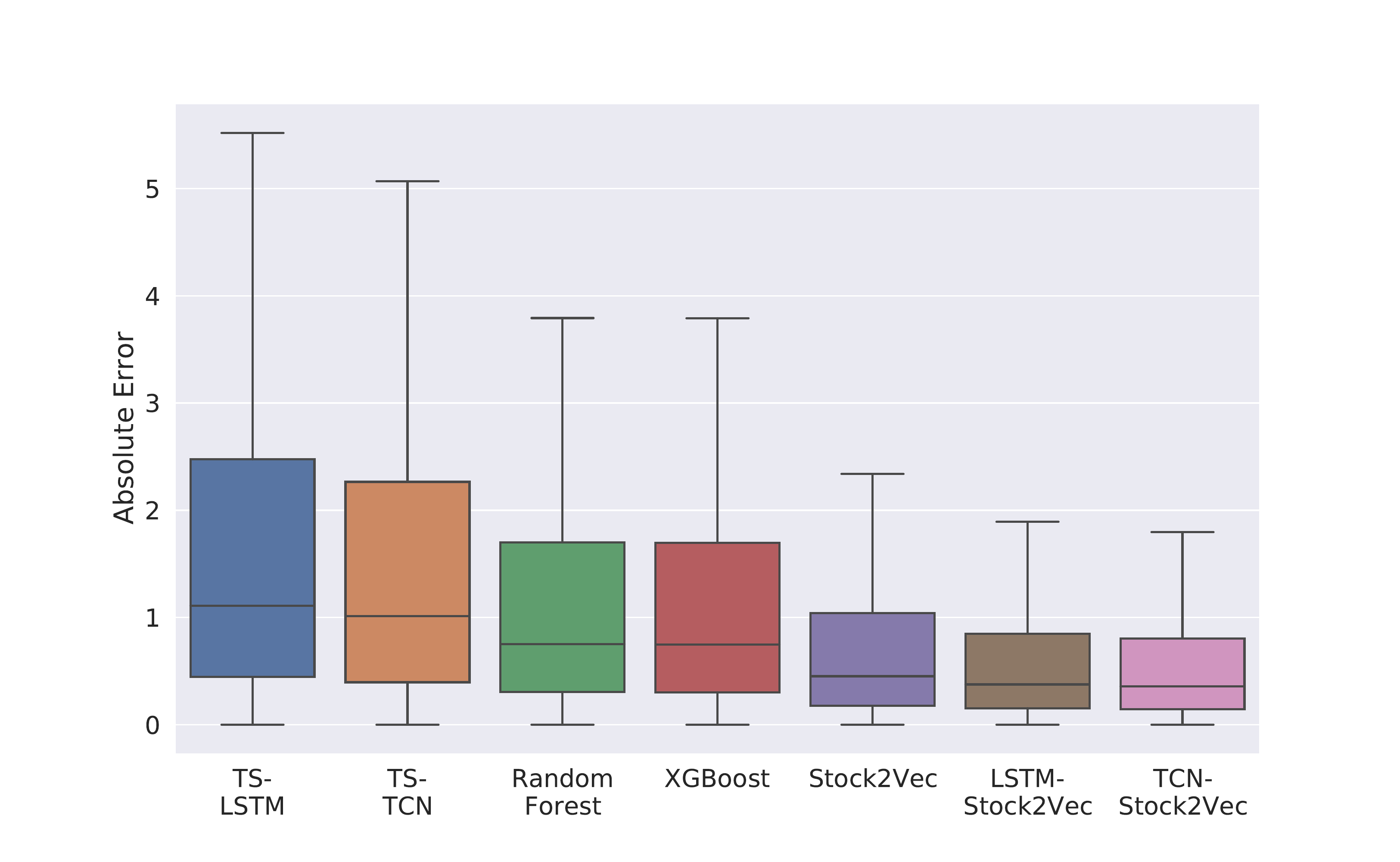}
    \caption{Boxplot comparison of the absolute prediction errors.}
    \label{fig:s2v_boxplot}
\end{figure}

Table \ref{tab:metrics_all} and Figure \ref{fig:s2v_boxplot} report the overall average (over the individual assets) forecasting performance of the out-of-sample period from 
2019-08-16 to 2020-02-14. 
We observe that TS-LSTM and TS-TCN perform worst. We can conlude that  this is because these two models only consider the target series and ignore all other features. TCN outperforms LSTM, probably since it is capable of extracting temporal patterns over long history without more effectively  gradient vanishing problem. Moreover, the training speed of our 18-layer TCN is about five times faster than that of LSTM per iteration (aka batch) with GPU, and the overall training speed (given all overhead included) is also around two to three times faster. 
With learning from all the features, the random forest and XGBoost models perform better than purely timeseries-based TS-LSTM and TS-TCN, with the XGBoost predictions are slightly better than that from random forest.
This demonstrates the usefulness of our data source, 
especially the external information combined into the inputs. We can then observe that
despite  having the same input as random forest and XGBoost, the proposed our Stock2Vec model further improves accuracy of the  predictions, as the RMSE, MAE, MAPE and RMSPE decrease by about 36\%, 38\%, 41\% and 43\% over the XGBoost predictions, respectively.
This indicates that the use of deep learning models, in particular the Stock2Vec embedding improves the predictions, by more effectively learning from the features over the tree-based ensemble models.
With integration of temporal modules, there is again a significant improvement of performance in terms of prediction accuracy. The two hybrid models LSTM-Stock2Vec and TCN-Stock2Vec not only learn from features we give explicitly, but also employ either a hidden state or a convolutional temporal feature mapping to implicitly learn relevant information from historical data. 
Our TCN-Stock2Vec achieves the best performance across all models, as the RMSE and MAE decreases by about 25\%, while the MAPE decreases by 20\% and the RMSPE decreases by 14\%, comparing with Stock2Vec without the temporal module.

\input{stock/tabs/MetricsAll}

Figure \ref{fig:s2v_boxplot} shows the boxplots of the prediction errors of different approaches, from which we can see our proposed models achieve smaller absolute prediction errors in terms of not only the mean also the variance, which indicates more robust forecast.
The median absolute prediction errors (and the interquartile range, i.e., IQR) of our TS-TCN model is around 1.01 (1.86), while they are around 0.74 (1.39),  0.45 (0.87), and 0.36 (0.66) for XGBoost, Stock2Vec and TCN-Stock2Vec, respectively. 

Similarly, we aggregate the metrics on the sector level, and calculate the average performance within each sector. We report the RMSE, MAE, MAPE, and RMSPE in Tables \ref{tab:sector_rmse}, \ref{tab:sector_mae}, \ref{tab:sector_mape}, and \ref{tab:sector_rmspe}, respectively, 
from which we can see again our Stock2Vec performs better than the two tree-ensemble models for all sectors, and adding the temporal module would further improve the forecasting accuracy. TCN-Stock2Vec achieves the best RMSE, MAE, MAPE and RMSPE in all sectors with one exception. Better performance on different aggregated levels demonstrates the power of our proposed models.

We further showcase the predicted results of 20 symbols to gauge the forecasting performance of our model under a wide range of industries, volatilities, growth patterns and other general conditions. 
The stocks have been chosen to evaluate how the proposed methodologies would perform under different circumstances. For instance, Amazon’s (AMZN) stock was consistently increasing in price across the analysis period, while the stock price of Verizon (VZ) was very stable, and Chevron’s stock (CVX) had both periods of growth and decline. In
addition, these 20 stocks captured several industries: (a) retail (e.g., Walmart), (b) restaurants (e.g., McDonald’s), (c) finance and banks (e.g., JPMorgan Chase and Goldman Sachs), (d) energy and oil \& gas (e.g., Chevron), (e) techonology (e.g., Facebook), (f) communications (e.g., Verizon), etc.
Table \ref{tab:symbol20_rmse}, \ref{tab:symbol20_mae}, \ref{tab:symbol20_mape}, \ref{tab:symbol20_rmape}  show the out-of-sample RMSE, MAE, MAPE and RMAPE, respectively, from the predictions given by the five models we discussed above.
Again, Stock2Vec generally performs better than random forest and XGBoost, and the two hybrid models have quite similar performance which is significantly better than that of others. While there also exist a few stocks on which LSTM-Stock2Vec or even Stock2Vec without temporal module produce most accurate predictions, for most of the stocks, TCN-Stock2Vec model performs the best.
This demonstrates our models generalize well to most symbols.

Furthermore, we plot the prediction pattern of the competing models for the abovementioned stocks on the test set in \ref{sec:s2v_plots21}, compared to the actual daily prices. We observe that the random forest and XGBoost models predict up-and-downs with a lag for most of the time, as the current price plays too much a role as a predictor, probably mainly due to the correct scaling reason. And there occasionally exist several flat predictions over a period for some stocks  (see approximately 2019/09 in Figure \ref{fig:plot_CVX}, 2020/01 in Figure \ref{fig:plot_FB}, and 2019/12 in Figure \ref{fig:plot_WMT}), which is a typical effect of tree-based methods, indicates insufficient splitting and underfitting despite so many ensemble trees were used.
With entity embeddings, our Stock2Vec model can learn from the features much more effectively,
its predictions coincide with the actual up-and downs much more accurately.
Although 
it overestimates the volatility by exaggerating the amplitude as well as the frequency of oscillations,
the overall prediction errors are getting smaller than the two tree-ensemble models.
And our LSTM-Stock2Vec and TCN-Stock2Vec models further benefit from the temporal learning modules by
automatically capturing the historical characteristics from time series data, especially the nonlinear trend and complex seasonality that are difficult to be captured by hand-engineered features such as technical indicators, as well as the common temporal factors that are shared among all series across the whole market. As a result, 
with the ability to extract the autoregressive dependencies over long term both within and across series from historical data, the predictions from these two models alleviate wild oscillations, 
and are much more close to the actual prices,
while still correctly predict the up-and-downs for most of the time with effective learning from input features.



%% file: stock/tabs/MetricsAll.tex
\begin{table}[bth!]
    \centering 
    \caption{Average performance comparison. }
    \footnotesize
    \begin{tabular}{|c||c|c|c|c|}    \hline
 & RMSE  & MAE  & MAPE(\%) & RMSPE (\%) \\ \hline \hline

TS-LSTM & 6.35 & 2.36 & 1.62 & 2.07 \\
TS-TCN & 5.79 & 2.15 & 1.50 & 1.96 \\ \hline 
Random Forest & 4.86 & 1.67 & 1.31 & 1.92 \\
XGBoost & 4.57 & 1.66 & 1.28 & 1.83 \\ \hline 
Stock2Vec & 2.94 & 1.04 & 0.76 & 1.05 \\ \hline 
LSTM-Stock2Vec & 2.57 & 0.85 & 0.68 & 1.04 \\
TCN-Stock2Vec & \textbf{2.22} & \textbf{0.78} & \textbf{0.61} & \textbf{0.90}

\\ \hline
    \end{tabular}
    \label{tab:metrics_all}

\end{table}

%% file: stock/conclusions.tex
Our argument that implicitly learning Alphas and Betas upon cross-sectional data from CAPM perspective is novel, however, it is more of an insight rather than systematic analysis.
In this paper, we built a global hybrid deep learning models to forecast the S\&P stock prices.  We applied the state-of-the-art 1-D dilated causal convolutional layers (TCN) to extract the temporal features from the historical information, which helps us to refine learning of the Alphas. In order to integrate the Beta information into the model, we learn a single model that learns from the data over the whole market, and applied entity embeddings for the categorical features, in particular, we obtained the Stock2Vec that reveals the relationship among stocks in the market, our model can be seen as supervised dimension reduction method in that point of view. 
The experimental results show our models improve the forecasting performance. Although not demonstrated in this work, learning a global model from the data over the entire market can give us an additional benefit that it can handle the cold-start problem, in which some series may contain very little data (i.e., many missing values), our model has the ability to infer the historical information with the structure learned from other series as well as the correlation between the cold-start series and the market. It might not be accurate, but is much informative than that learned from little data in the single series.


    
    
There are several other directions that we can dive deeper as the future work. 
First of all, the stock prices are heavily affected by  external  information, combining extensive crowd-sourcing, social media and financial news data may facilitate a better understanding of collective human behavior on the market, which could help the effective decision making for investors. These data can be obtained from the internet, we  could expand the data source and combine their influence in the model as extra features. In addition, although we have shown that the convolutional layers have several advantages over the most widely used recurrent neural  network  layers  for  time  series,  the  temporal  learning  layers  in  our  model  could  be replaced by any other type, for instance, the recent advances of attention models could be a  good  candidate.  
Also, 
more sophisticated models can be adopted to build Stock2Vec, by keeping the goal in mind that we aim at learning the implicit intrinsic relationship between stock series.
In addition, learning the relationship over the market would be helpful for us to build portfolio aiming at maximizing the investment gain, e.g., by using standard Markowitz portfolio optimization to find the positions. In that case, simulation of trading in the market should provide us more realistic and robust performance evaluation than those aggregated levels we reported above. Liquidity and market impacts can be taken into account in the simulation, and we can use Profit \& Loss (P\&L) and the Sharpe ratio as the evaluation metrics.

%% file: stock/tabs/sectorRMSE.tex
\begin{table}[!h]
    \caption{Sector level RMSE comparison}
    \centering
    \label{tab:sector_rmse}
\resizebox{.8\textwidth}{!}{%
    \begin{tabular}{|c||c|c|c|c|c|} \hline
 & Random Forest & XGBoost & Stock2Vec & LSTM-Stock2Vec & TCN-Stock2Vec \\ \hline \hline 

Basic Industries & 1.70 & 1.61 & 1.06 & 0.85 & \textbf{0.76} \\
Capital Goods & 11.46 & 10.30 & 6.25 & 6.01 & \textbf{5.10} \\
Consumer Durables & 1.78 & 1.67 & 0.99 & 0.93 & \textbf{0.83} \\
Consumer Non-Durables & 1.57 & 1.55 & 0.98 & 0.87 & \textbf{0.75} \\
Consumer Services & 4.75 & 4.69 & 3.34 & 2.76 & \textbf{2.30} \\
Energy & 1.50 & 1.44 & 0.76 & 0.76 & \textbf{0.67} \\
Finance & 2.08 & 2.06 & 1.39 & 1.05 & \textbf{1.00} \\
HealthCare & 3.44 & 3.37 & 1.95 & 1.98 & \textbf{1.60} \\
Miscellaneous & 8.23 & 7.96 & 5.22 & 4.14 & \textbf{3.73} \\
Public Utilities & 0.94 & 0.95 & 0.64 & \textbf{0.52} & 0.52 \\
Technology & 4.20 & 4.23 & 2.91 & 1.90 & \textbf{1.94} \\
Transportation & 2.00 & 1.90 & 1.15 & 1.03 & \textbf{0.88}

\\ \hline 
    \end{tabular}
}
\end{table}

%% file: stock/tabs/sectorMAE.tex
\begin{table}[!h]
    \caption{Sector level MAE 
    comparison}
    \centering
    \label{tab:sector_mae}
\resizebox{.8\textwidth}{!}{%
    \begin{tabular}{|c||c|c|c|c|c|} \hline
& Random Forest & XGBoost & Stock2Vec & LSTM-Stock2Vec & TCN-Stock2Vec \\ \hline \hline

Basic Industries & 1.06 & 1.03 & 0.64 & 0.52 & \textbf{0.49} \\
Capital Goods & 3.13 & 3.07 & 1.93 & 1.57 & \textbf{1.47} \\
Consumer Durables & 1.21 & 1.18 & 0.71 & 0.63 & \textbf{0.57} \\
Consumer Non-Durables & 0.96 & 0.93 & 0.57 & 0.52 & \textbf{0.45} \\
Consumer Services & 1.83 & 1.84 & 1.19 & 0.98 & \textbf{0.88} \\
Energy & 0.98 & 0.95 & 0.50 & 0.51 & \textbf{0.45} \\
Finance & 1.19 & 1.17 & 0.79 & 0.55 & \textbf{0.54} \\
HealthCare & 1.99 & 1.96 & 1.15 & 1.10 & \textbf{0.92} \\
Miscellaneous & 3.18 & 3.18 & 2.08 & 1.56 & \textbf{1.44} \\
Public Utilities & 0.63 & 0.64 & 0.44 & \textbf{0.33} & 0.34 \\
Technology & 1.95 & 1.98 & 1.26 & 0.92 & \textbf{0.91} \\
Transportation & 1.26 & 1.23 & 0.74 & 0.65 & \textbf{0.56}

\\ \hline
    \end{tabular}
}
\end{table}

%% file: stock/tabs/sectorMAPE.tex
\begin{table}[!h]
    \caption{Sector level MAPE (\%) comparison}
    \centering
    \label{tab:sector_mape}
\resizebox{.8\textwidth}{!}{%
    \begin{tabular}{|c||c|c|c|c|c|} \hline
& Random Forest & XGBoost & Stock2Vec & LSTM-Stock2Vec & TCN-Stock2Vec \\ \hline \hline 

Basic Industries & 1.34 & 1.31 & 0.74 & 0.65 & \textbf{0.61} \\
Capital Goods & 1.21 & 1.24 & 0.76 & 0.59 & \textbf{0.56} \\
Consumer Durables & 1.30 & 1.26 & 0.73 & 0.68 & \textbf{0.60} \\
Consumer Non-Durables & 1.48 & 1.32 & 0.76 & 0.85 & \textbf{0.65} \\
Consumer Services & 1.24 & 1.23 & 0.71 & 0.66 & \textbf{0.59} \\
Energy & 2.04 & 1.88 & 0.97 & 1.08 & \textbf{0.92} \\
Finance & 1.18 & 1.16 & 0.74 & \textbf{0.53} & 0.53 \\
HealthCare & 1.43 & 1.35 & 0.79 & 0.79 & \textbf{0.65} \\
Miscellaneous & 1.23 & 1.23 & 0.81 & 0.66 & \textbf{0.60} \\
Public Utilities & 0.88 & 0.90 & 0.57 & 0.49 & \textbf{0.47} \\
Technology & 1.44 & 1.43 & 0.83 & 0.68 & \textbf{0.66} \\
Transportation & 1.26 & 1.23 & 0.71 & 0.66 & \textbf{0.57}

\\ \hline 
    \end{tabular}
}
\end{table}

%% file: stock/tabs/sectorRMSPE.tex
\begin{table}[!ht]
    \caption{Sector level RMSPE (\%) comparison}
    \centering
    \label{tab:sector_rmspe}
\resizebox{.8\textwidth}{!}{%
    \begin{tabular}{|c||c|c|c|c|c|} \hline
 & Random Forest & XGBoost & Stock2Vec & LSTM-Stock2Vec & TCN-Stock2Vec \\ \hline \hline 

Basic Industries & 1.86 & 1.80 & 0.98 & 0.91 & \textbf{0.83} \\
Capital Goods & 1.63 & 1.65 & 1.01 & 0.83 & \textbf{0.75} \\
Consumer Durables & 1.79 & 1.68 & 0.96 & 0.95 & \textbf{0.81} \\
Consumer Non-Durables & 2.41 & 2.02 & 1.13 & 1.37 & \textbf{1.01} \\
Consumer Services & 1.88 & 1.82 & 0.99 & 1.07 & \textbf{0.91} \\
Energy & 2.89 & 2.66 & 1.29 & 1.51 & \textbf{1.25} \\
Finance & 1.60 & 1.56 & 1.00 & 0.78 & \textbf{0.72} \\
HealthCare & 2.17 & 2.00 & 1.15 & 1.24 & \textbf{0.99} \\
Miscellaneous & 1.66 & 1.63 & 1.05 & 0.95 & \textbf{0.81} \\
Public Utilities & 1.25 & 1.23 & 0.74 & 0.71 & \textbf{0.64} \\
Technology & 2.09 & 2.00 & 1.13 & 1.04 & \textbf{0.95} \\
Transportation & 1.76 & 1.68 & 0.98 & 0.95 & \textbf{0.80}

\\ \hline 
    \end{tabular}
}
\end{table}

%% file: stock/tabs/s20RMSE.tex
\begin{table}[h!]
    \caption{RMSE comparison of different models for the one-day ahead forecasting on different symbols}
    \label{tab:symbol20_rmse}
    \centering
\resizebox{.8\textwidth}{!}{%
    \begin{tabular}{|c||c|c|c|c|c|} \hline
 & Random Forest & XGBoost & Stock2Vec & LSTM-Stock2Vec & TCN-Stock2Vec \\ \hline \hline

AAPL (Apple) & 4.71 & 4.52 & 2.86 & 2.16 & \textbf{1.81} \\
AFL (Aflac) & 0.59 & 0.62 & 0.46 & \textbf{0.31} & \textbf{0.27} \\
AMZN (Amazon.com) & 29.91 & 28.47 & 23.80 & 17.73 & \textbf{14.45} \\
BA (Boeing) & 6.00 & 6.44 & 3.98 & 3.83 & \textbf{3.49} \\
CVX (Chevron) & 1.42 & 1.62 & 1.03 & 0.75 & \textbf{0.65} \\
DAL (Delta Air Lines) & 0.79 & 0.77 & 0.48 & 0.40 & \textbf{0.32} \\
DIS (Walt Disney) & 1.95 & 1.91 & 1.17 & 1.10 & \textbf{0.92} \\
FB (Facebook) & 3.51 & 5.54 & 2.15 & 1.72 & \textbf{1.44} \\
GE (General Electric) & 0.39 & 0.30 & \textbf{0.14} & 0.29 & 0.18 \\
GM (General Motors) & 0.58 & 0.57 & 0.30 & 0.30 & \textbf{0.28} \\
GS (Goldman Sachs Group) & 3.11 & 3.00 & 1.86 & \textbf{1.27} & 1.31 \\
JNJ (Johnson \& Johnson) & 1.80 & 1.49 & 1.00 & 0.93 & \textbf{0.80} \\
JPM (JPMorgan Chase) & 1.72 & 1.63 & 1.59 & \textbf{0.66} & 0.68 \\
MAR (Marriott Int'l) & 2.02 & 2.02 & 1.52 & \textbf{0.89} & 1.07 \\
KO (Coca-Cola) & 0.49 & 0.50 & 0.32 & 0.26 & \textbf{0.25} \\
MCD (McDonald's) & 2.67 & 2.50 & 1.51 & 1.26 & \textbf{1.16} \\
NKE (Nike) & 1.27 & 1.23 & 1.01 & \textbf{0.61} & 0.62 \\
PG (Procter \& Gamble) & 1.43 & 1.35 & 0.91 & 0.70 & \textbf{0.61} \\
VZ (Verizon Communications) & 0.54 & 0.55 & 0.46 & 0.29 & \textbf{0.26} \\
WMT (Walmart) & 1.34 & 1.43 & 1.06 & 0.55 & \textbf{0.50}

\\ \hline 
\end{tabular}
}
\end{table}

%% file: stock/tabs/s20MAE.tex
\begin{table}[h!]
    \caption{MAE comparison of different models for the one-day ahead forecasting on different symbols}
    \label{tab:symbol20_mae}
    \centering
\resizebox{.8\textwidth}{!}{%
    \begin{tabular}{|c||c|c|c|c|c|} \hline
 & Random Forest & XGBoost & Stock2Vec & LSTM-Stock2Vec & TCN-Stock2Vec \\ \hline \hline

AAPL (Apple) & 3.63 & 3.56 & 2.15 & 1.72 & \textbf{1.42} \\
AFL (Aflac) & 0.45 & 0.44 & 0.35 & \textbf{0.20} & 0.21 \\
AMZN (Amazon.com) & 22.19 & 21.36 & 17.87 & 11.53 & \textbf{10.29} \\
BA (Boeing) & 4.59 & 5.10 & 2.87 & 2.87 & \textbf{2.74} \\
CVX (Chevron) & 1.07 & 1.22 & 0.75 & 0.57 & \textbf{0.50} \\
DAL (Delta Air Lines) & 0.59 & 0.58 & 0.36 & 0.29 & \textbf{0.24} \\
DIS (Walt Disney) & 1.37 & 1.40 & 0.87 & 0.77 & \textbf{0.67} \\
FB (Facebook) & 2.54 & 3.80 & 1.65 & 1.16 & \textbf{1.06} \\
GE (General Electric) & 0.30 & 0.22 & \textbf{0.11} & 0.25 & 0.15 \\
GM (General Motors) & 0.44 & 0.44 & 0.23 & 0.23 & \textbf{0.22} \\
GS (Goldman Sachs Group) & 2.48 & 2.37 & 1.31 & \textbf{1.01} & 1.05 \\
JNJ (Johnson \& Johnson) & 1.21 & 1.04 & 0.72 & 0.64 & \textbf{0.59} \\
JPM (JPMorgan Chase) & 1.34 & 1.23 & 1.17 & \textbf{0.51} & 0.52 \\
MAR (Marriott Int'l) & 1.63 & 1.66 & 1.13 & \textbf{0.65} & 0.87 \\
KO (Coca-Cola) & 0.39 & 0.37 & 0.25 & 0.19 & \textbf{0.19} \\
MCD (McDonald's) & 1.99 & 1.96 & 1.26 & 0.89 & \textbf{0.89} \\
NKE (Nike) & 0.97 & 0.98 & 0.77 & \textbf{0.46} & 0.49 \\
PG (Procter \& Gamble) & 1.14 & 1.03 & 0.70 & 0.52 & \textbf{0.48} \\
VZ (Verizon Communications) & 0.43 & 0.42 & 0.36 & 0.22 & \textbf{0.20} \\
WMT (Walmart) & 1.02 & 1.10 & 0.87 & \textbf{0.41} & 0.41

\\ \hline 
\end{tabular}
}
\end{table}

%% file: stock/tabs/s20MAPE.tex
\begin{table}[h!]
    \caption{MAPE (\%) comparison of different models for the one-day ahead forecasting on different symbols}
    \label{tab:symbol20_mape}
    \centering
\resizebox{.8\textwidth}{!}{%
    \begin{tabular}{|c||c|c|c|c|c|} \hline
 & Random Forest & XGBoost & Stock2Vec & LSTM-Stock2Vec & TCN-Stock2Vec \\ \hline \hline

AAPL (Apple) & 1.43 & 1.39 & 0.80 & 0.68 & \textbf{0.54}  \\
AFL (Aflac) & 0.88 & 0.86 & 0.66 & \textbf{0.39}  & 0.39 \\
AMZN (Amazon.com) & 1.21 & 1.17 & 0.97 & 0.63 & \textbf{0.56}  \\
BA (Boeing) & 1.33 & 1.47 & 0.82 & 0.83 & \textbf{0.80}  \\
CVX (Chevron) & 0.94 & 1.06 & 0.65 & 0.50 & \textbf{0.43}  \\
DAL (Delta Air Lines) & 1.03 & 1.02 & 0.63 & 0.51 & \textbf{0.43}  \\
DIS (Walt Disney) & 0.99 & 1.01 & 0.61 & 0.55 & \textbf{0.48}  \\
FB (Facebook) & 1.29 & 1.92 & 0.82 & 0.59 & \textbf{0.54}  \\
GE (General Electric) & 2.99 & 2.13 & \textbf{1.10}  & 2.53 & 1.44 \\
GM (General Motors) & 1.22 & 1.23 & 0.63 & 0.63 & \textbf{0.61}  \\
GS (Goldman Sachs Group) & 1.14 & 1.09 & 0.59 & \textbf{0.46}  & 0.48 \\
JNJ (Johnson \& Johnson) & 0.90 & 0.77 & 0.51 & 0.47 & \textbf{0.43}  \\
JPM (JPMorgan Chase) & 1.08 & 1.00 & 0.90 & \textbf{0.40}  & 0.42 \\
MAR (Marriott Int'l) & 1.21 & 1.23 & 0.81 & \textbf{0.48}  & 0.63 \\
KO (Coca-Cola) & 0.72 & 0.68 & 0.45 & \textbf{0.35}  & 0.35 \\
MCD (McDonald's) & 0.98 & 0.96 & 0.61 & \textbf{0.44}  & 0.44 \\
NKE (Nike) & 1.05 & 1.06 & 0.80 & \textbf{0.49}  & 0.53 \\
PG (Procter \& Gamble) & 0.94 & 0.85 & 0.57 & 0.43 & \textbf{0.40}  \\
VZ (Verizon Communications) & 0.73 & 0.71 & 0.60 & 0.37 & \textbf{0.34}  \\
WMT (Walmart) & 0.88 & 0.94 & 0.73 & \textbf{0.35} & 0.35

\\ \hline 
\end{tabular}
}
\end{table}

%% file: stock/tabs/s20RMSPE.tex
\begin{table}[h!]
    \caption{RMAPE (\%) comparison of different models for the one-day ahead forecasting on different symbols}
    \label{tab:symbol20_rmape}
    \centering
\resizebox{.8\textwidth}{!}{%
    \begin{tabular}{|c||c|c|c|c|c|} \hline
 & Random Forest & XGBoost & Stock2Vec & LSTM-Stock2Vec & TCN-Stock2Vec \\ \hline \hline

AAPL (Apple) & 1.89 & 1.76 & 1.04 & 0.85 & \textbf{0.68} \\
AFL (Aflac) & 1.15 & 1.19 & 0.87 & 0.60 & \textbf{0.53} \\
AMZN (Amazon.com) & 1.60 & 1.55 & 1.28 & 0.95 & \textbf{0.78} \\
BA (Boeing) & 1.74 & 1.85 & 1.13 & 1.11 & \textbf{1.02} \\
CVX (Chevron) & 1.25 & 1.42 & 0.88 & 0.65 & \textbf{0.57} \\
DAL (Delta Air Lines) & 1.39 & 1.36 & 0.83 & 0.71 & \textbf{0.57} \\
DIS (Walt Disney) & 1.41 & 1.38 & 0.81 & 0.79 & \textbf{0.66} \\
FB (Facebook) & 1.77 & 2.75 & 1.06 & 0.85 & \textbf{0.73} \\
GE (General Electric) & 3.96 & 2.89 & \textbf{1.35} & 2.91 & 1.72 \\
GM (General Motors) & 1.62 & 1.60 & 0.82 & 0.84 & \textbf{0.77} \\
GS (Goldman Sachs Group) & 1.44 & 1.39 & 0.84 & \textbf{0.58} & 0.61 \\
JNJ (Johnson \& Johnson) & 1.33 & 1.11 & 0.72 & 0.70 & \textbf{0.60} \\
JPM (JPMorgan Chase) & 1.40 & 1.33 & 1.20 & \textbf{0.53} & 0.54 \\
MAR (Marriott Int'l) & 1.49 & 1.50 & 1.07 & \textbf{0.66} & 0.78 \\
KO (Coca-Cola) & 0.90 & 0.92 & 0.57 & 0.47 & \textbf{0.45} \\
MCD (McDonald's) & 1.30 & 1.22 & 0.73 & 0.62 & \textbf{0.57} \\
NKE (Nike) & 1.38 & 1.34 & 1.03 & \textbf{0.65} & 0.67 \\
PG (Procter \& Gamble) & 1.19 & 1.11 & 0.73 & 0.58 & \textbf{0.50} \\
VZ (Verizon Communications) & 0.93 & 0.93 & 0.76 & 0.49 & \textbf{0.45} \\
WMT (Walmart) & 1.15 & 1.23 & 0.89 & 0.47 & \textbf{0.43}

\\ \hline 
\end{tabular}
}
\end{table}

%% file: stock/plots21.tex
\begin{figure}[!htb] 
	\centering 
		 \includegraphics[width=\textwidth]{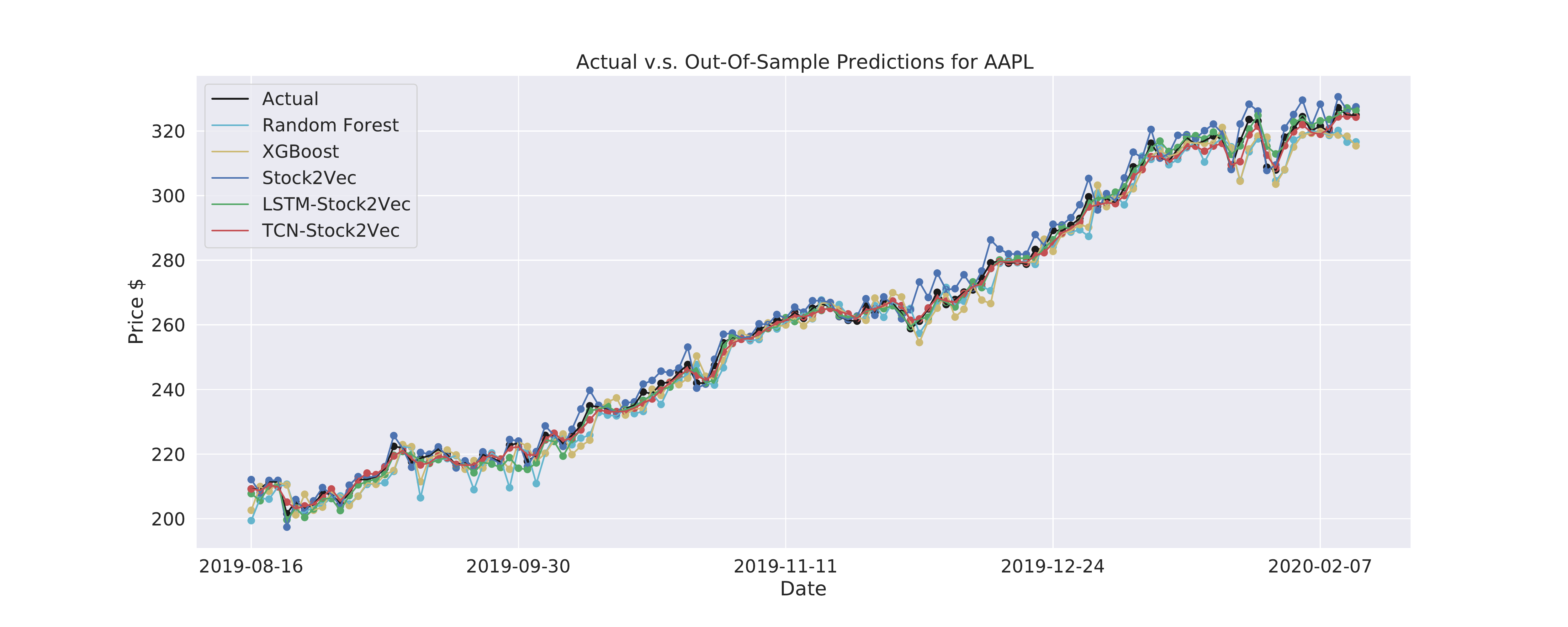}
	\caption{Showcase AAPL of predicted v.s. actual daily prices of one stock over test period, 2019/08/16-2020/02/14.}
	\label{fig:plot_AAPL}
\end{figure}

\begin{figure}[!htb] 
	\centering 
		 \includegraphics[width=\textwidth]{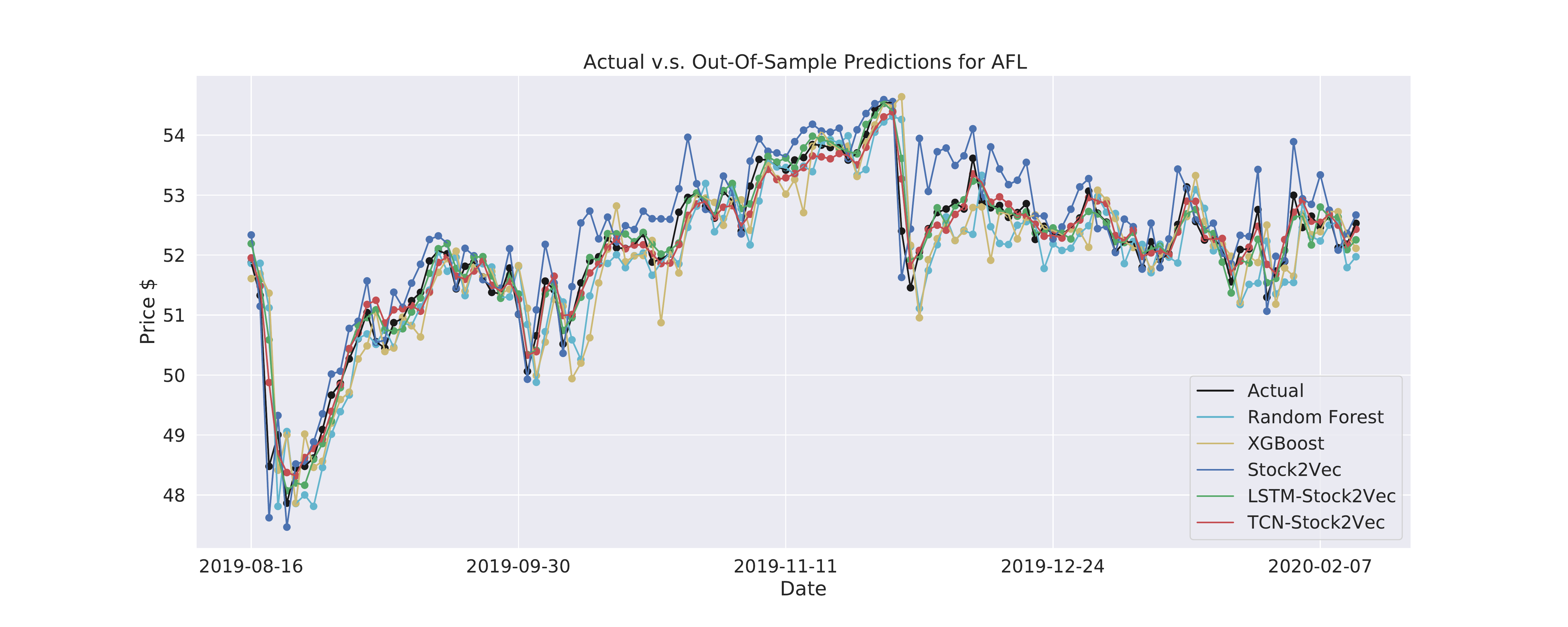}
	\caption{Showcase AFL of predicted v.s. actual daily prices of one stock over test period, 2019/08/16-2020/02/14.}
	\label{fig:plot_AFL}
\end{figure}

\begin{figure}[!htb] 
	\centering 
		 \includegraphics[width=\textwidth]{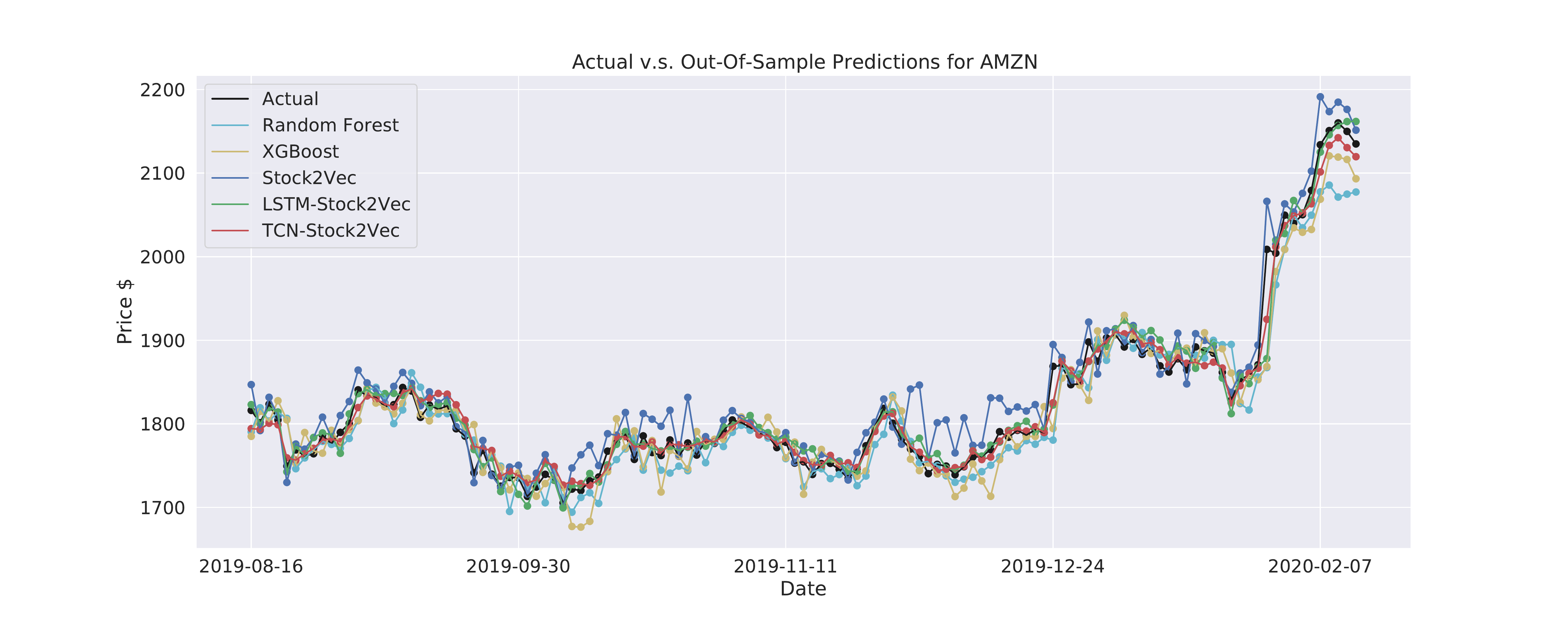}
	\caption{Showcase AMZN of predicted v.s. actual daily prices of one stock over test period, 2019/08/16-2020/02/14.}
	\label{fig:plot_AMZN}
\end{figure}

\begin{figure}[!htb] 
	\centering 
		 \includegraphics[width=\textwidth]{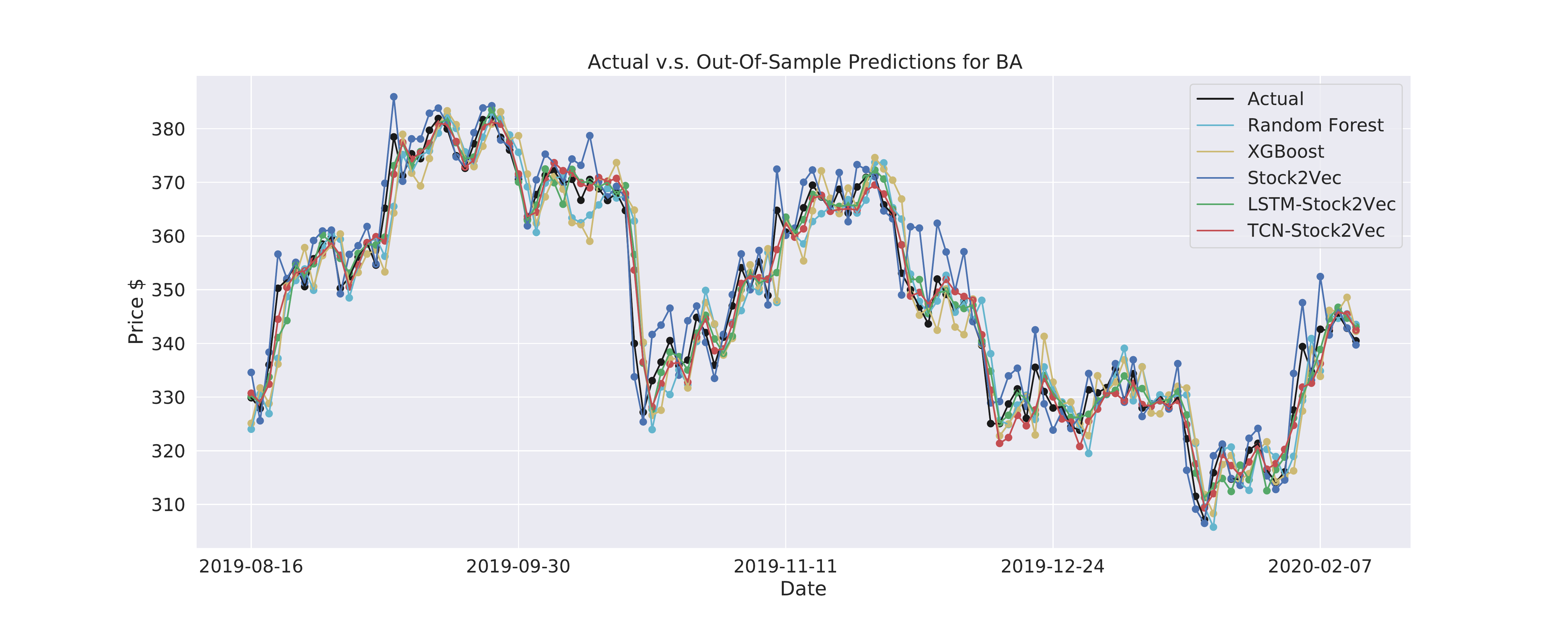}
	\caption{Showcase BA of predicted v.s. actual daily prices of one stock over test period, 2019/08/16-2020/02/14.}
	\label{fig:plot_BA}
\end{figure}

\begin{figure}[!htb] 
	\centering 
		 \includegraphics[width=\textwidth]{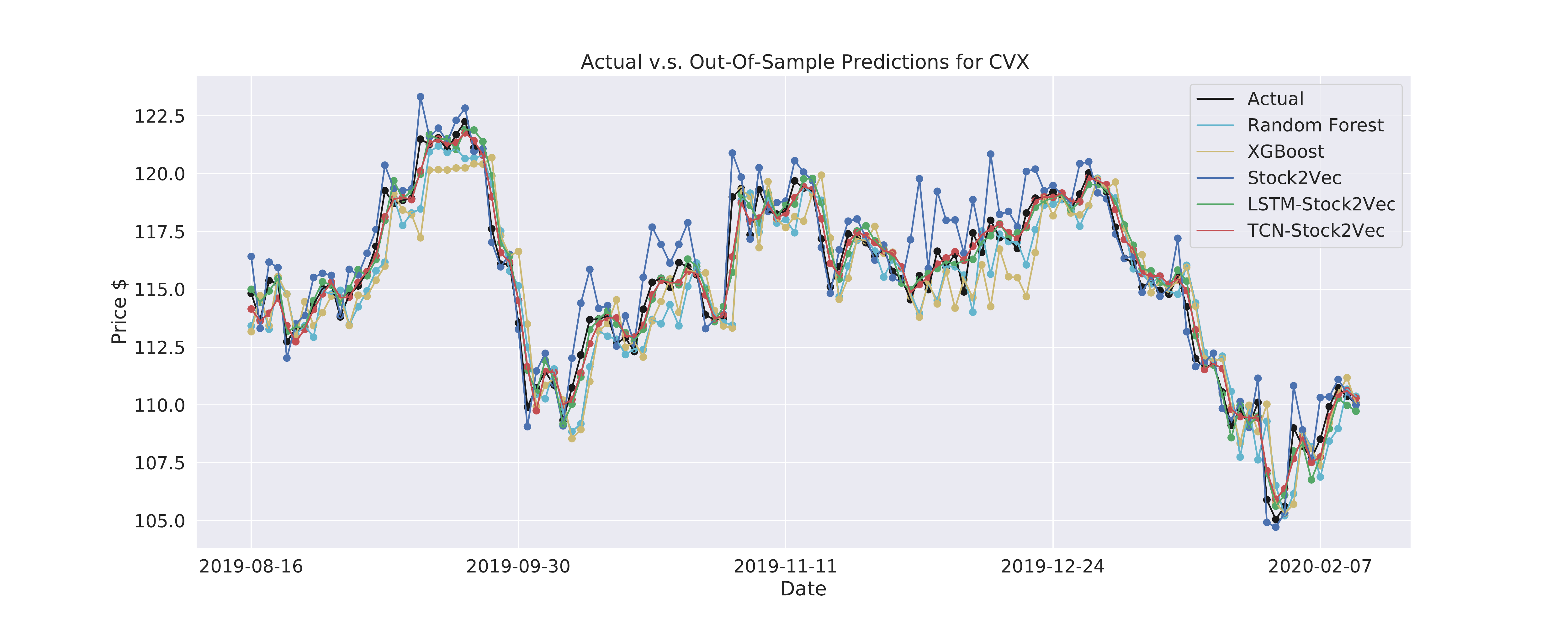}
	\caption{Showcase CVX of predicted v.s. actual daily prices of one stock over test period, 2019/08/16-2020/02/14.}
	\label{fig:plot_CVX}
\end{figure}

\begin{figure}[!htb] 
	\centering 
		 \includegraphics[width=\textwidth]{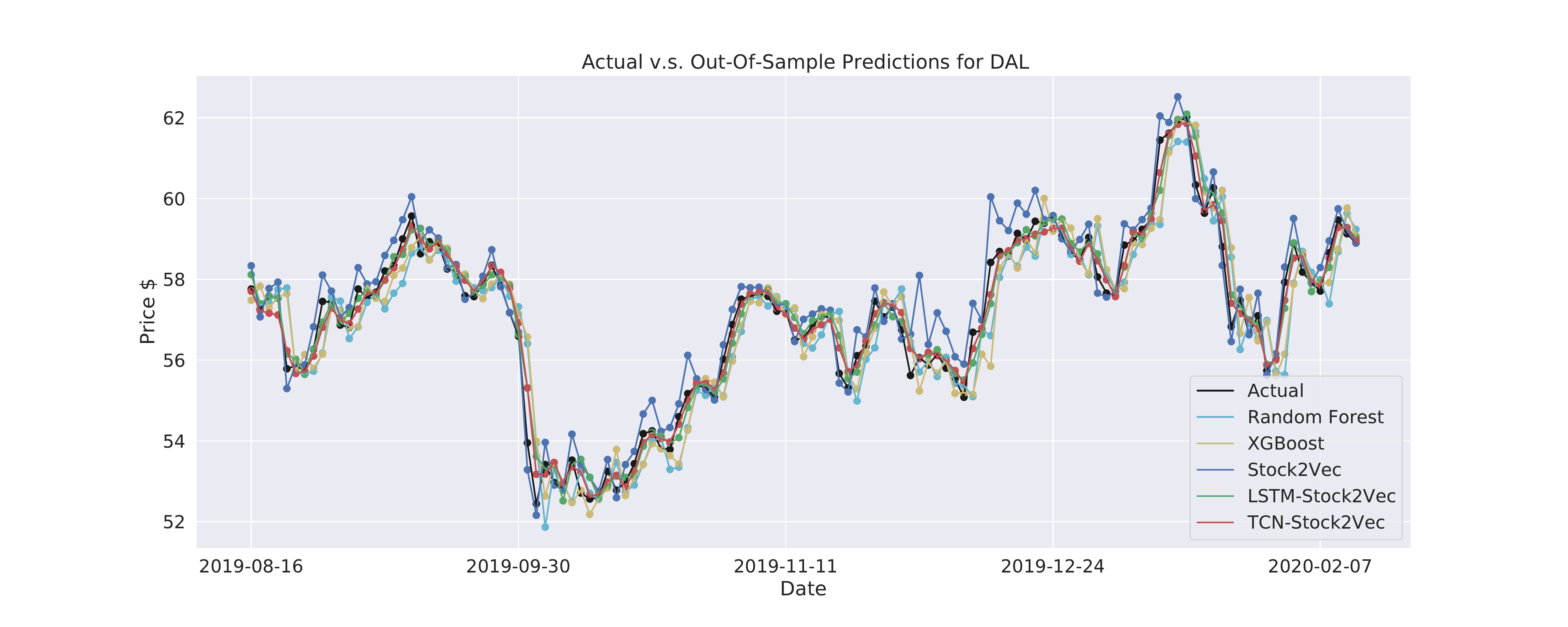}
	\caption{Showcase DAL of predicted v.s. actual daily prices of one stock over test period, 2019/08/16-2020/02/14.}
	\label{fig:plot_DAL}
\end{figure}

\begin{figure}[!htb] 
	\centering 
		 \includegraphics[width=\textwidth]{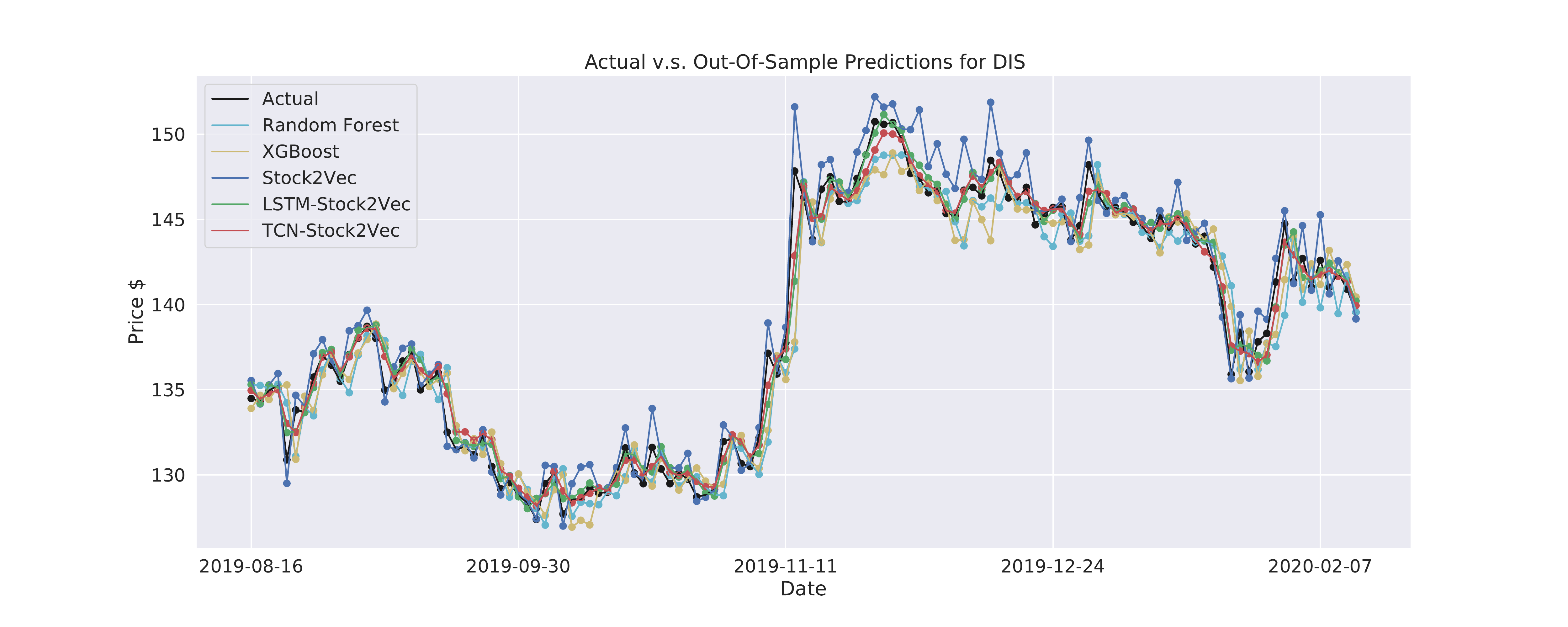}
	\caption{Showcase DIS of predicted v.s. actual daily prices of one stock over test period, 2019/08/16-2020/02/14.}
	\label{fig:plot_DIS}
\end{figure}

\begin{figure}[!htb] 
	\centering 
		 \includegraphics[width=\textwidth]{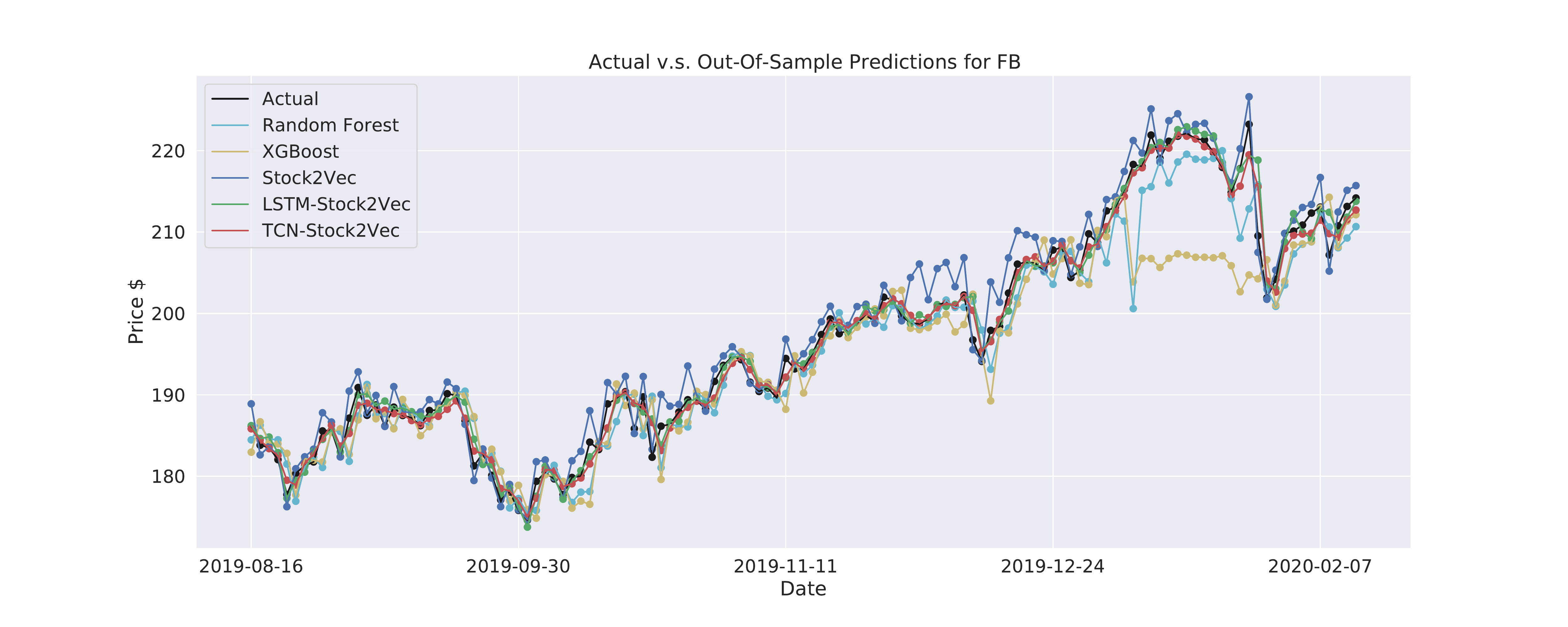}
	\caption{Showcase FB of predicted v.s. actual daily prices of one stock over test period, 2019/08/16-2020/02/14.}
	\label{fig:plot_FB}
\end{figure}

\begin{figure}[!htb] 
	\centering 
		 \includegraphics[width=\textwidth]{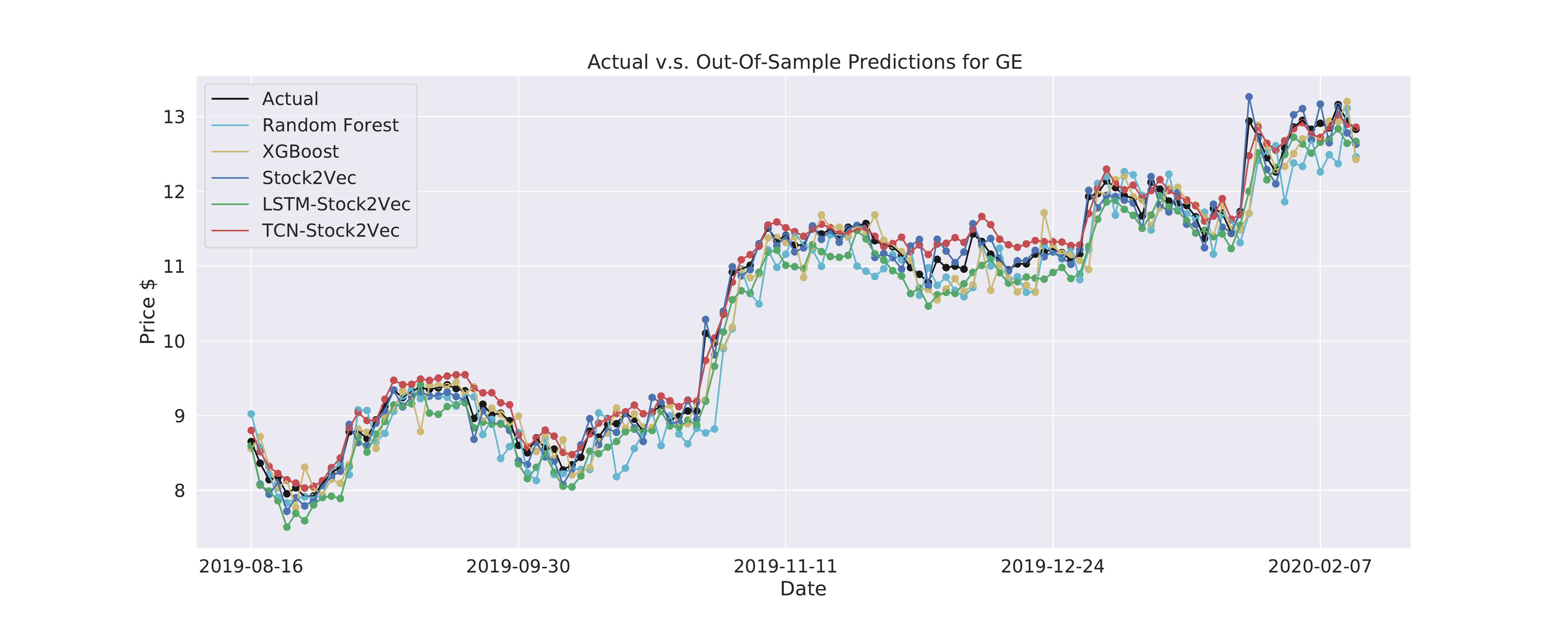}
	\caption{Showcase GE of predicted v.s. actual daily prices of one stock over test period, 2019/08/16-2020/02/14.}
	\label{fig:plot_GE}
\end{figure}

\begin{figure}[!htb] 
	\centering 
		 \includegraphics[width=\textwidth]{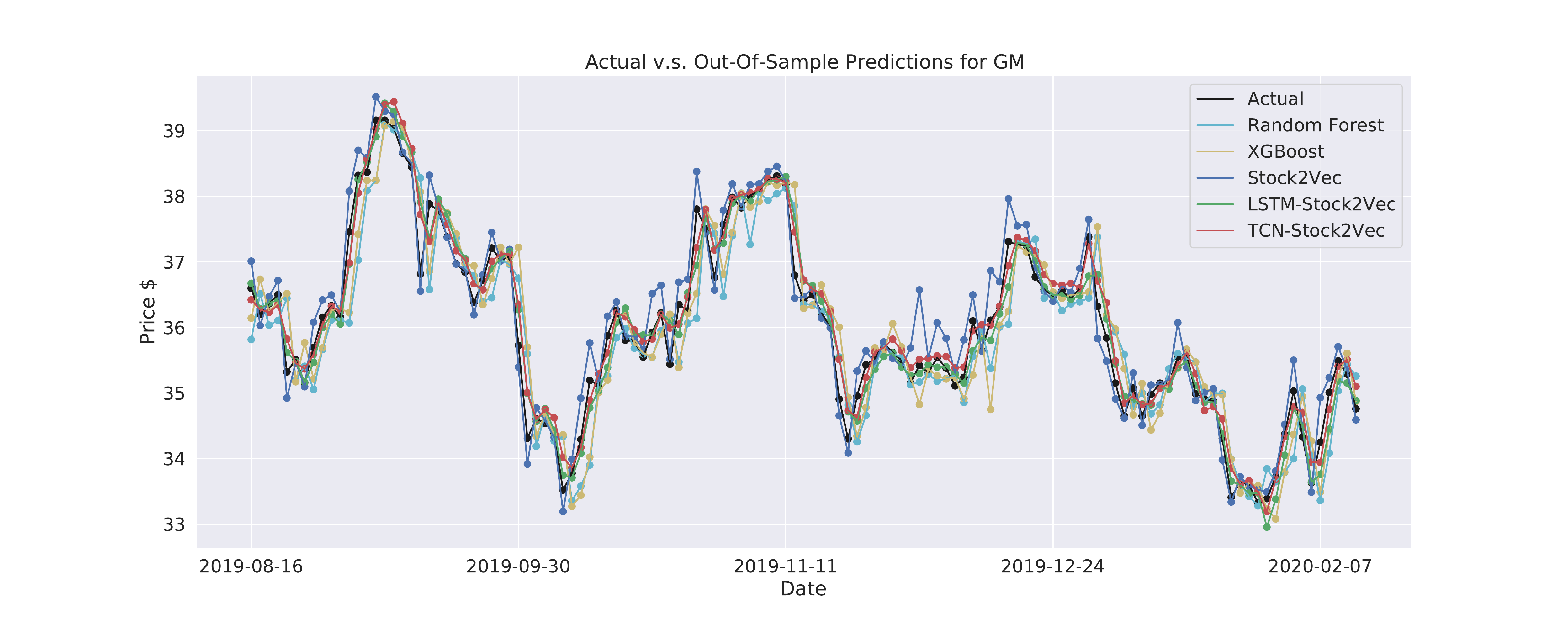}
	\caption{Showcase GM of predicted v.s. actual daily prices of one stock over test period, 2019/08/16-2020/02/14.}
	\label{fig:plot_GM}
\end{figure}


\begin{figure}[!htb] 
	\centering 
		 \includegraphics[width=\textwidth]{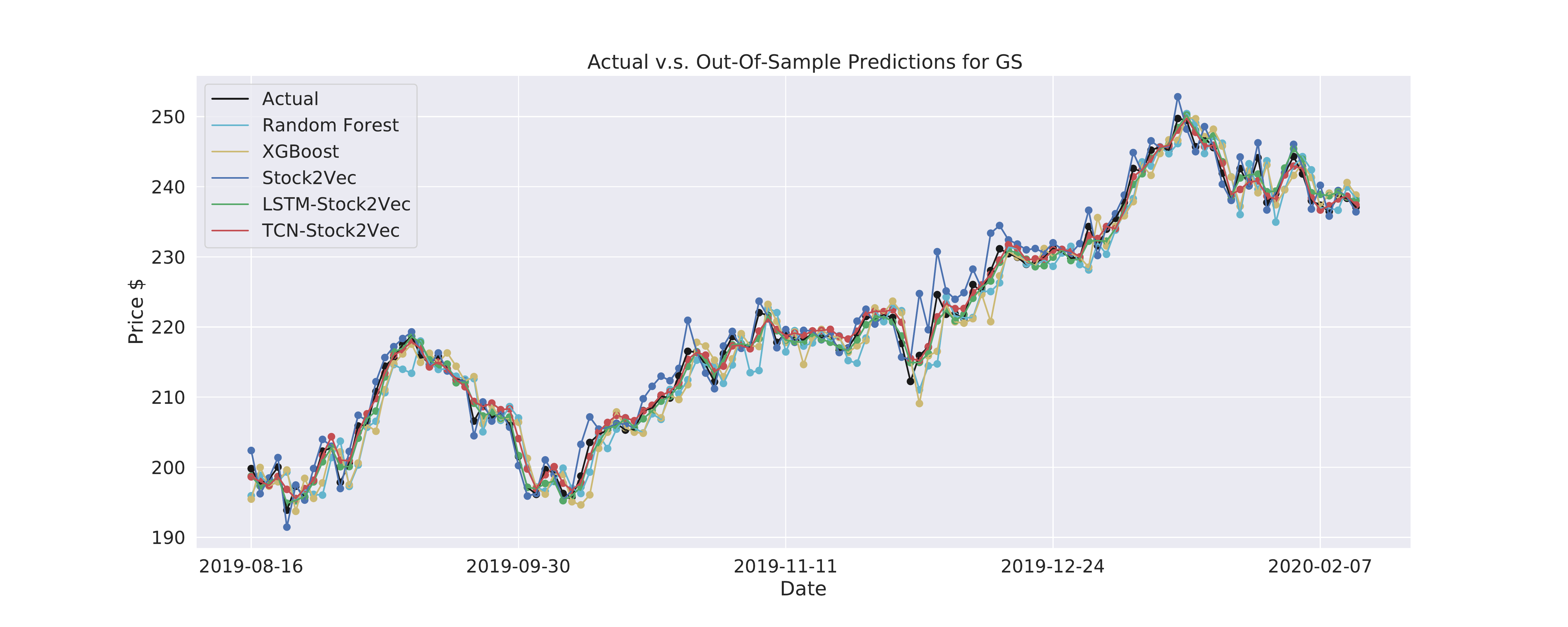}
	\caption{Showcase GS of predicted v.s. actual daily prices of one stock over test period, 2019/08/16-2020/02/14.}
	\label{fig:plot_GS}
\end{figure}

\begin{figure}[!htb] 
	\centering 
		 \includegraphics[width=\textwidth]{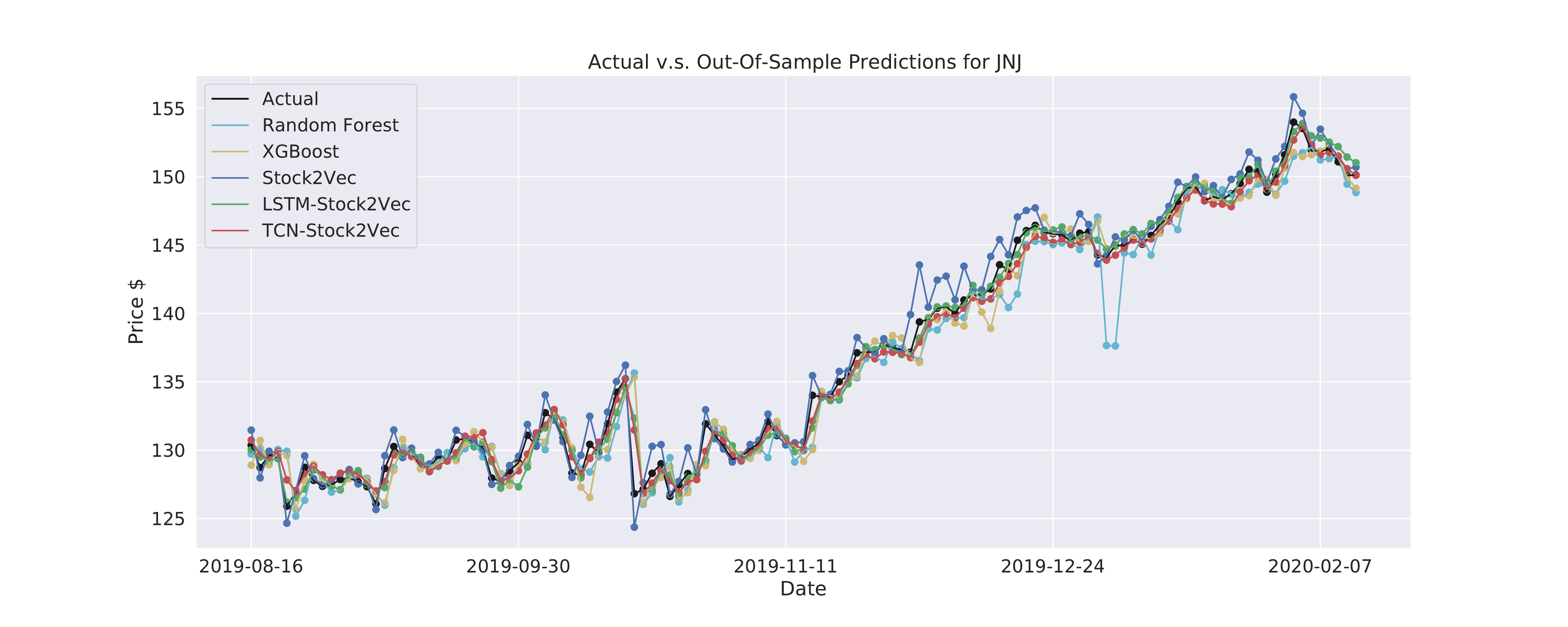}
	\caption{Showcase JNJ of predicted v.s. actual daily prices of one stock over test period, 2019/08/16-2020/02/14.}
	\label{fig:plot_JNJ}
\end{figure}

\begin{figure}[!htb] 
	\centering 
		 \includegraphics[width=\textwidth]{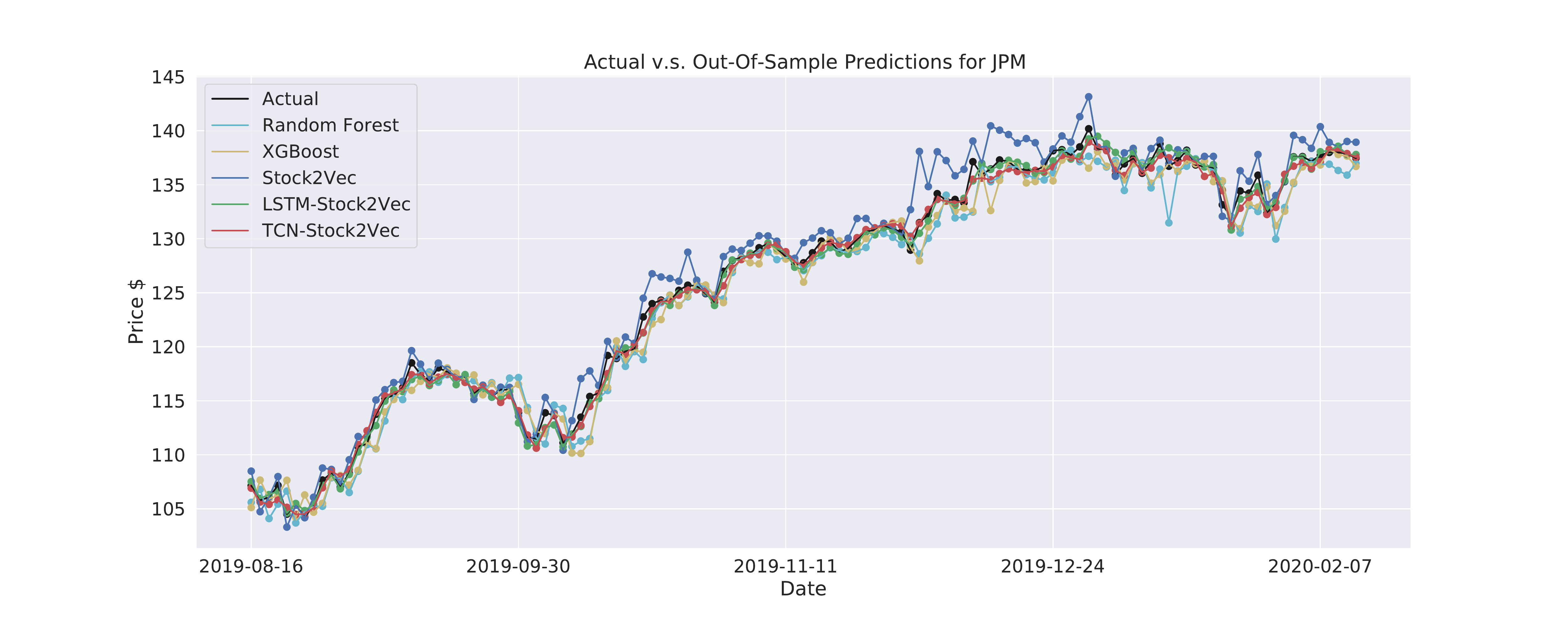}
	\caption{Showcase JPM of predicted v.s. actual daily prices of one stock over test period, 2019/08/16-2020/02/14.}
	\label{fig:plot_JPM}
\end{figure}

\begin{figure}[!htb] 
	\centering 
		 \includegraphics[width=\textwidth]{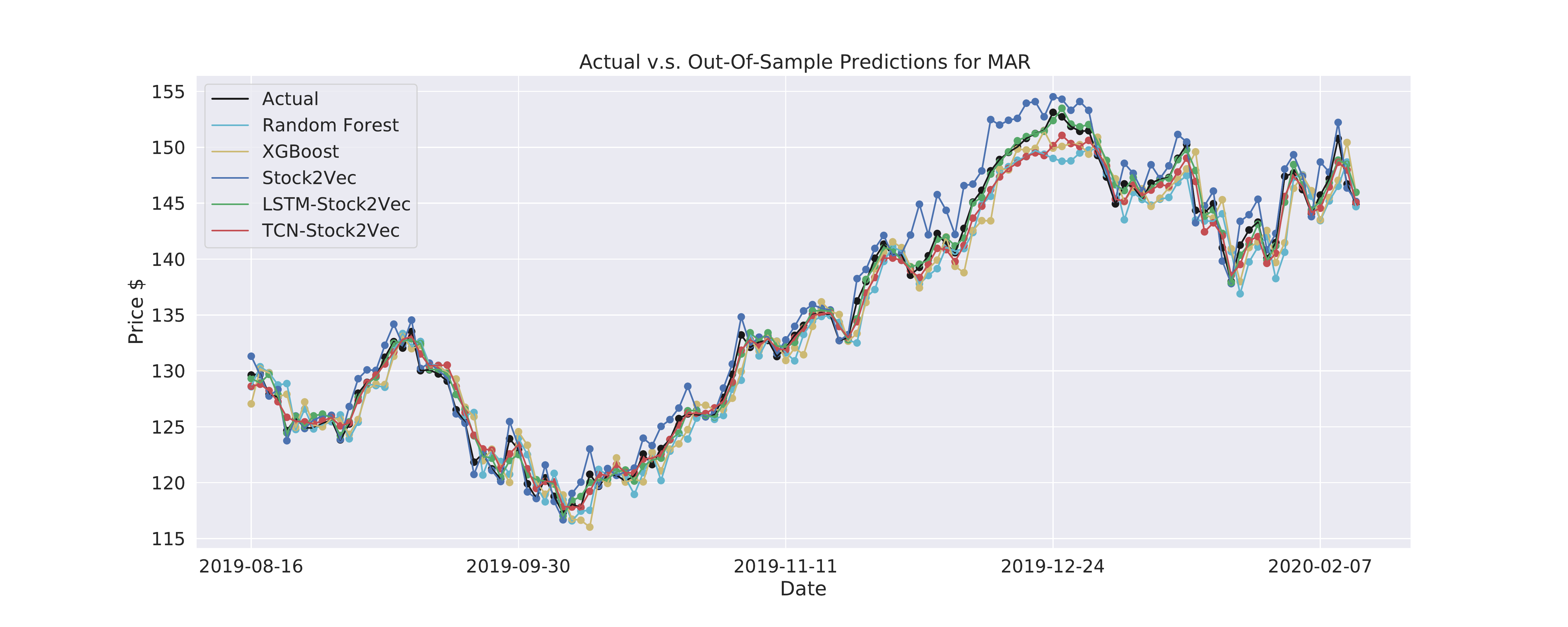}
	\caption{Showcase MAR of predicted v.s. actual daily prices of one stock over test period, 2019/08/16-2020/02/14.}
	\label{fig:plot_MAR}
\end{figure}

\begin{figure}[!htb] 
	\centering 
		 \includegraphics[width=\textwidth]{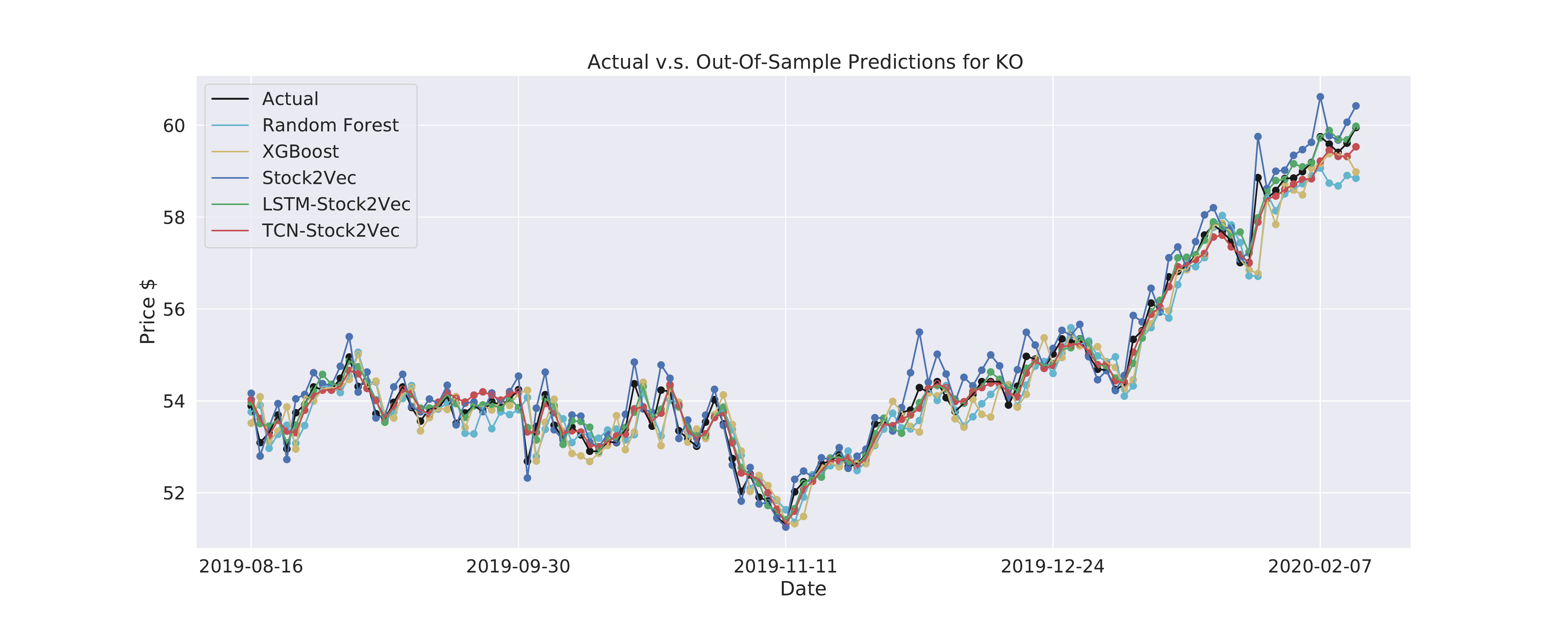}
	\caption{Showcase KO of predicted v.s. actual daily prices of one stock over test period, 2019/08/16-2020/02/14.}
	\label{fig:plot_KO}
\end{figure}

\begin{figure}[!htb] 
	\centering 
		 \includegraphics[width=\textwidth]{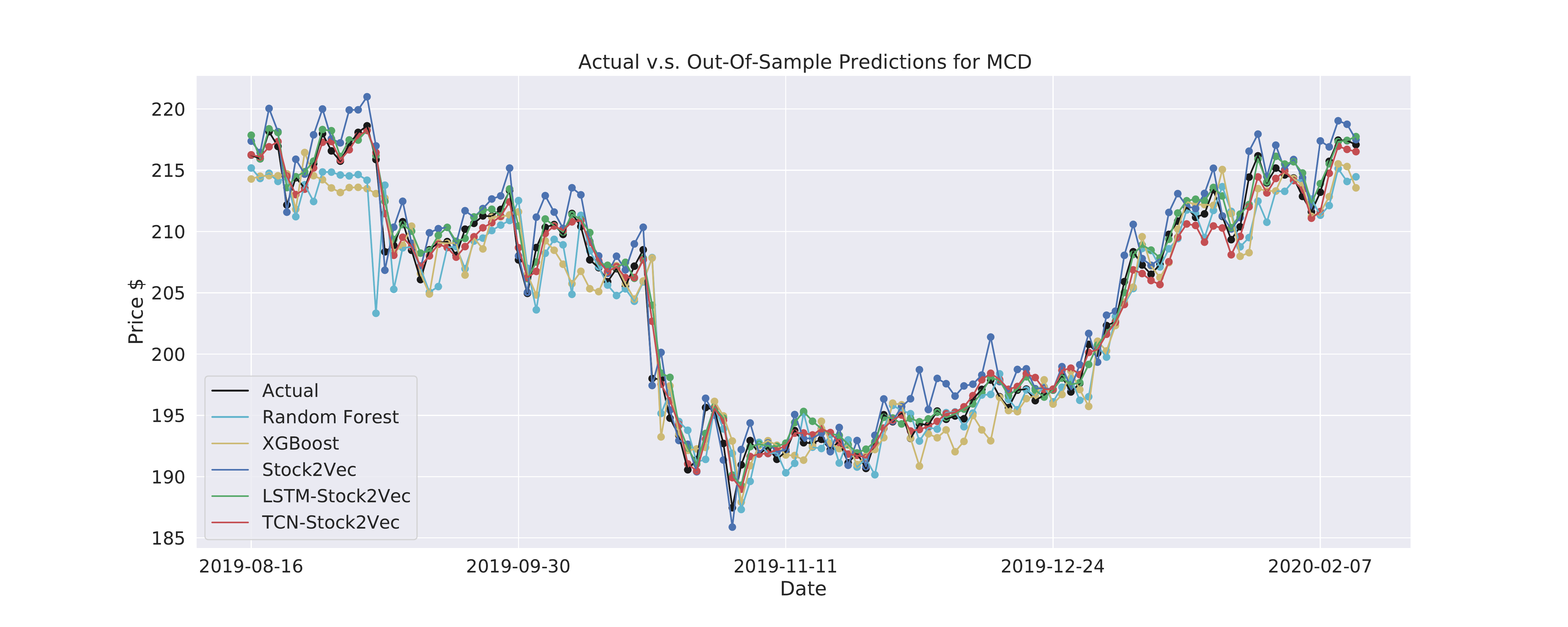}
	\caption{Showcase MCD of predicted v.s. actual daily prices of one stock over test period, 2019/08/16-2020/02/14.}
	\label{fig:plot_MCD}
\end{figure}

\begin{figure}[!htb] 
	\centering 
		 \includegraphics[width=\textwidth]{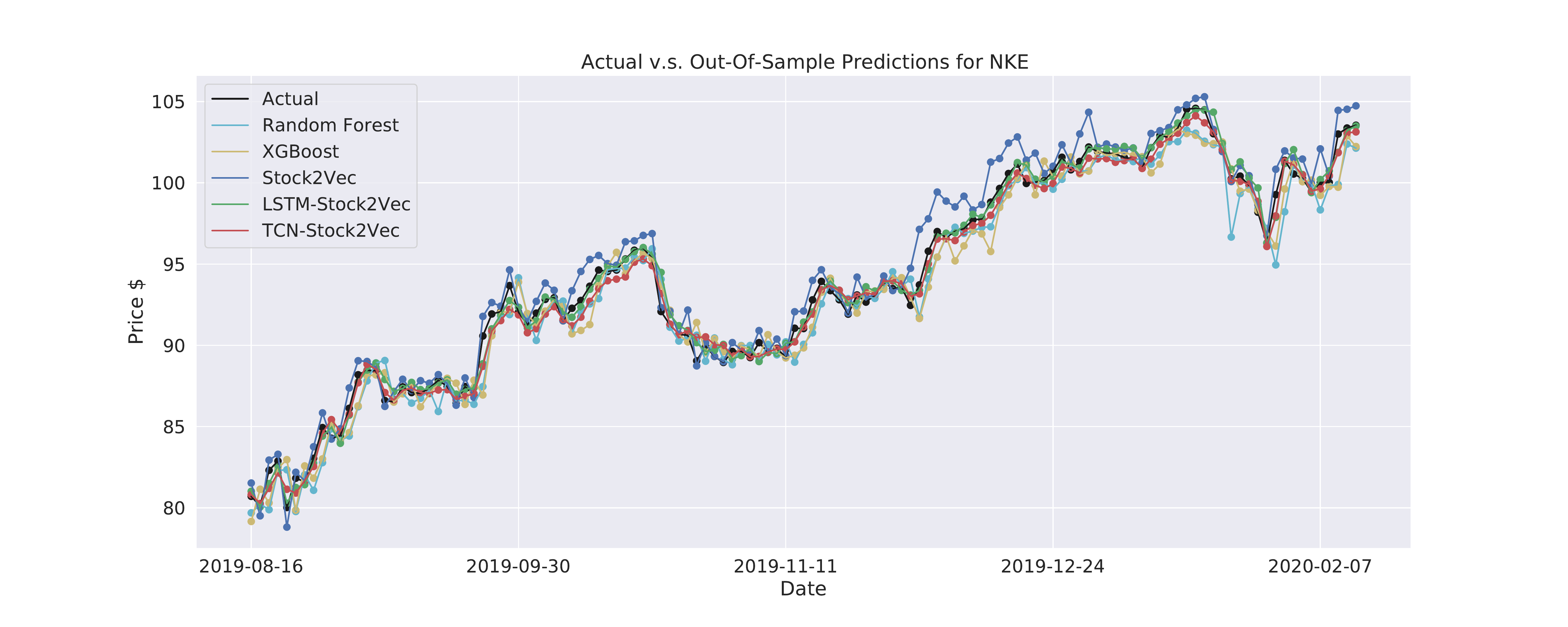}
	\caption{Showcase NKE of predicted v.s. actual daily prices of one stock over test period, 2019/08/16-2020/02/14.}
	\label{fig:plot_NKE}
\end{figure}

\begin{figure}[!htb] 
	\centering 
		 \includegraphics[width=\textwidth]{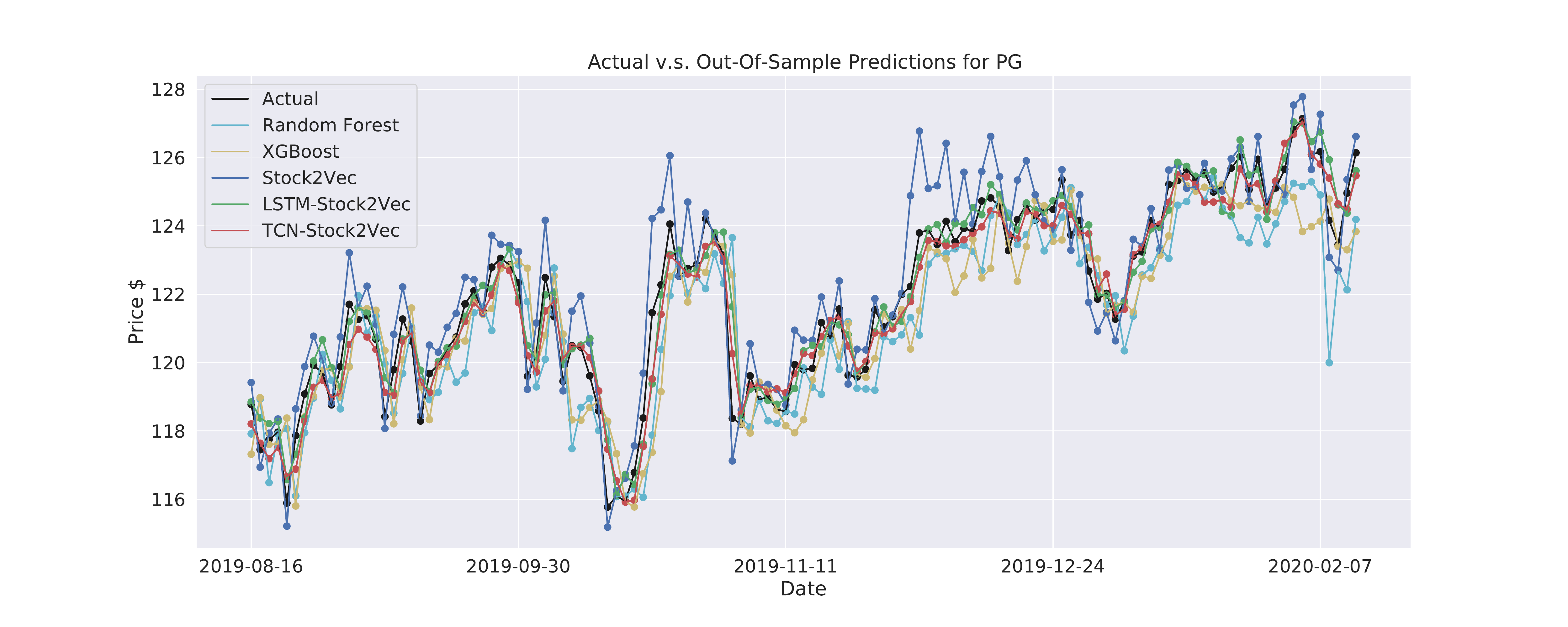}
	\caption{Showcase PG of predicted v.s. actual daily prices of one stock over test period, 2019/08/16-2020/02/14.}
	\label{fig:plot_PG}
\end{figure}

\begin{figure}[!htb] 
	\centering 
		 \includegraphics[width=\textwidth]{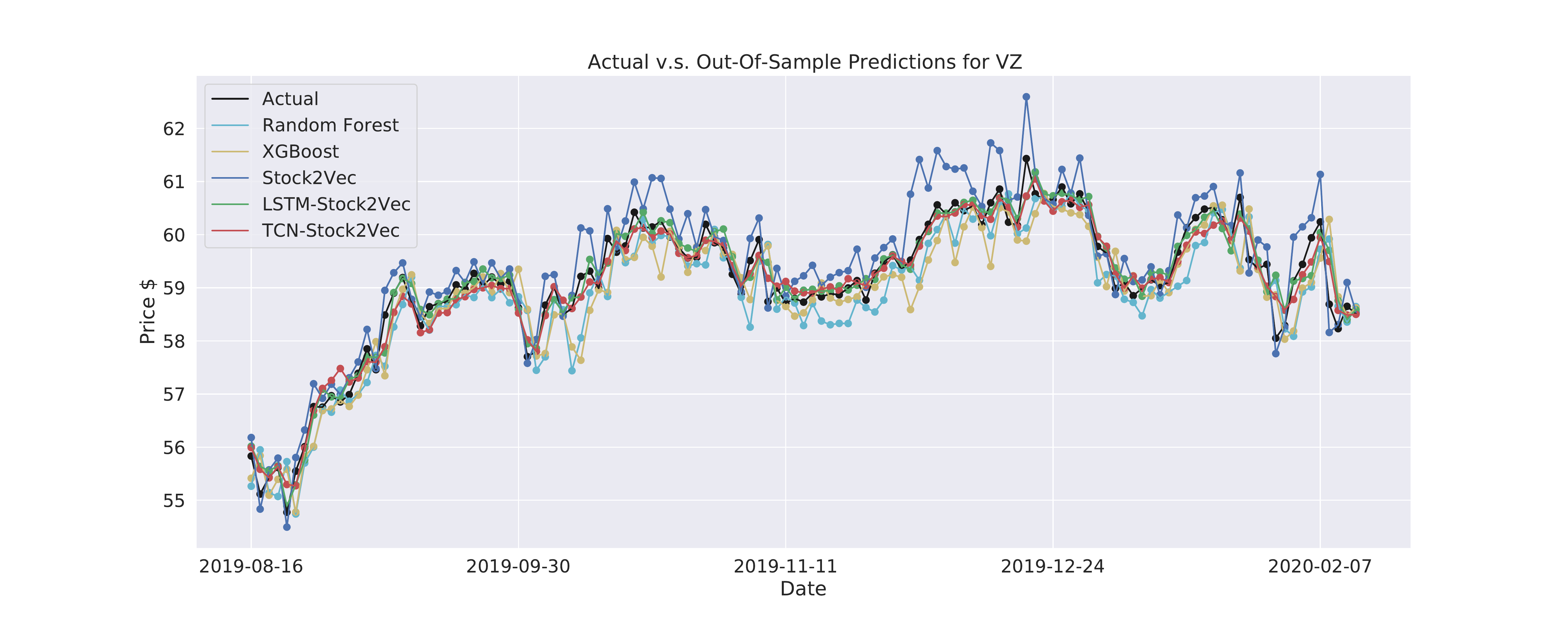}
	\caption{Showcase VZ of predicted v.s. actual daily prices of one stock over test period, 2019/08/16-2020/02/14.}
	\label{fig:plot_VZ}
\end{figure}

\begin{figure}[!htb] 
	\centering 
		 \includegraphics[width=\textwidth]{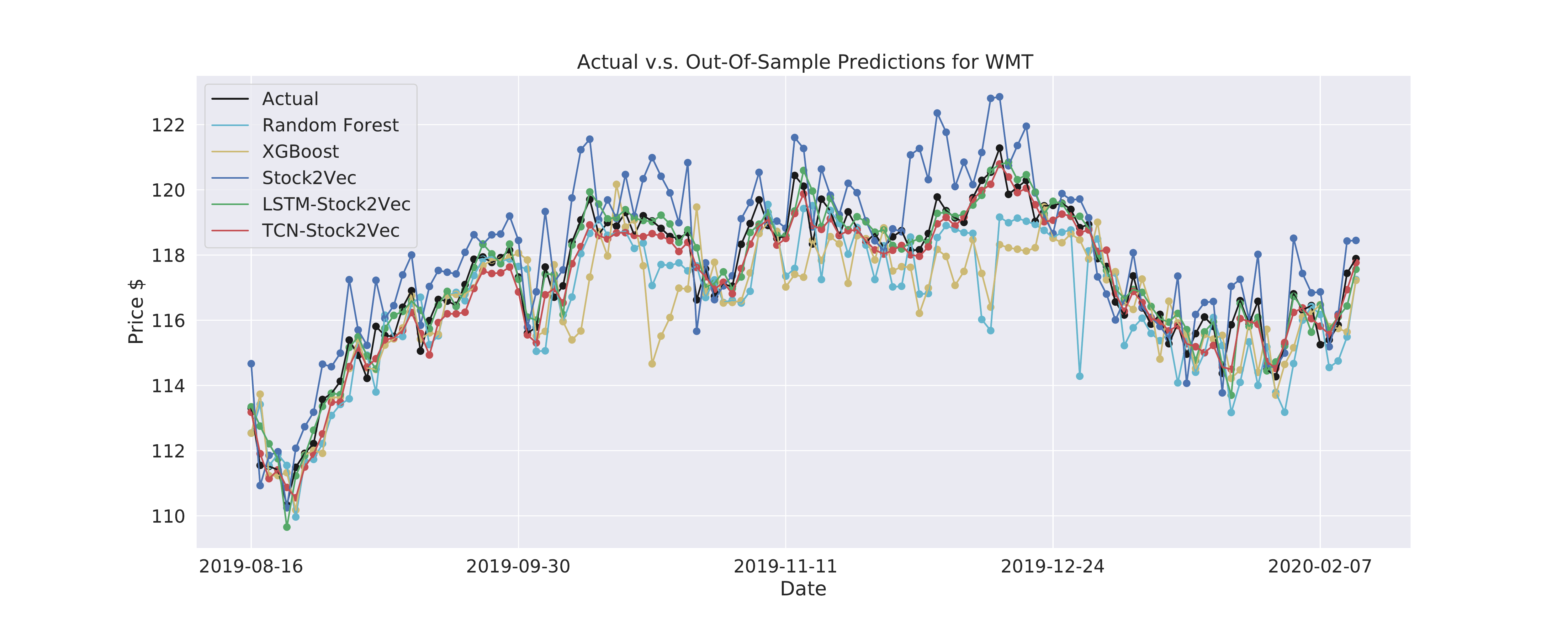}
	\caption{Showcase WMT of predicted v.s. actual daily prices of one stock over test period, 2019/08/16-2020/02/14.}
	\label{fig:plot_WMT}
\end{figure}

